\documentclass[fleqn,usenatbib]{mnras}

\usepackage{ulem}

\usepackage[T1]{fontenc}
\usepackage{ae,aecompl}

\usepackage{graphicx}    
\usepackage{amsmath}    

\usepackage{mathptmx}
\usepackage{txfonts}

\newcommand{\units}[1]{~\mathrm{#1}}
\newcommand{\msun}{\mathrm{M}_\odot}

\newcommand{\mvir}{\mathrm{M}_{200}}

\newcommand{\lx}{$L_{\mathrm{X}}$}
\newcommand{\lxmvir}{$L_{\mathrm{X}}-M_{\mathrm{vir}}$}
\newcommand{\lxmstar}{$L_{\mathrm{X}}-M_{\mathrm{star}}$}

\title[Origin of simulated X-ray coronae]{The origin of X-ray coronae around
    simulated disc galaxies}

\author[A. J. Kelly, A. Jenkins \& C. S. Frenk]{Ashley J.
    Kelly,$^{1}$\thanks{E-mail: a.j.kelly@durham.ac.uk} Adrian Jenkins,$^{1}$
    Carlos S. Frenk$^{1}$\\
$^{1}$Institute for Computational Cosmology, Department of Physics, Durham
University, Durham DH1 3LE, U.K\\}

\date{Accepted XXX. Received YYY; in original form ZZZ}

\pubyear{2020}

\begin{document}\label{firstpage}
\pagerange{\pageref{firstpage}--\pageref{lastpage}}
\maketitle

\begin{abstract}
The existence of hot, accreted gaseous coronae around massive galaxies is a long-standing central prediction of galaxy formation models in the $\Lambda$CDM cosmology. While observations now confirm that extraplanar hot gas is present around late-type galaxies, the origin of the gas is uncertain with suggestions that galactic feedback could be the dominant source of energy powering the emission.  We investigate the origin and X-ray properties of the hot gas that surrounds galaxies of halo mass, $(10^{11}-10^{14}) \mathrm{M}_\odot$, in the cosmological hydrodynamical EAGLE simulations. We find that the central X-ray emission, $\leq 0.10 R_{\mathrm{vir}}$, of halos of mass $\leq 10^{13} \mathrm{M}_\odot$ originates from gas heated by supernovae (SNe). However, beyond this region, a quasi-hydrostatic, accreted atmosphere dominates the X-ray emission in halos of mass $\geq 10^{12} \mathrm{M}_\odot$. We predict that a dependence on halo mass of the hot gas to dark matter mass fraction can significantly change the slope of the $L_{\mathrm{X}}-M_{\mathrm{vir}}$ relation (which is typically assumed to be $4/3$ for clusters) and we derive the scaling law appropriate to this case. As the gas fraction in halos increases with halo mass, we find a steeper slope for the $L_{\mathrm{X}}-M_{\mathrm{vir}}$ in lower mass halos, $\leq 10^{14} \mathrm{M}_\odot$. This varying gas fraction is driven by active galactic nuclei (AGN) feedback. We also identify the physical origin of the so-called ``missing feedback'' problem, the apparently low X-ray luminosities observed from high star-forming, low-mass galaxies. This is explained by the ejection of SNe-heated gas from the central regions of the halo.
\end{abstract}


\section{Introduction}\label{intro}

A long-standing fundamental prediction of galaxy formation theories
within the $\Lambda$CDM cosmological framework is that a significant
fraction of the baryons in massive dark matter halos should reside
in a hot atmosphere that surrounds the central galaxy
\citep{white:1991}. However, the limited detections of significant
extraplanar X-ray emission around MW-mass galaxies challenge these
models.

In the early galaxy formation models of \cite{white:1978} and
\cite{white:1991} gas is accreted from the intergalactic medium (IGM)
and falls into a dark matter potential. The subsequent behaviour of
the accreting gas depends on the `cooling radius', which is the radius
at which the cooling time of the gas is equal to the dynamical time of
the halo. In low mass halos, the cooling radius extends well beyond
the halo and, consequently, if inflowing gas is shock-heated,
it can efficiently cool and rapidly accrete onto the central galaxy on
a timescale comparable to the free-fall time of the halo. However, in
halos of virial mass,  $M_{\mathrm{vir}} \gtrapprox 10^{12} \units{\msun}$,
the cooling radius lies deep within the halo. Thus, infalling gas
shock-heats to the virial temperature of the halo and settles into a hot,
quasi-hydrostatic atmosphere of gas (\citealt{larson:1974}; the idea of an extended,
hot gas corona around the Milky Way was already suggested by
\citealt{spitzer:1956}). In the innermost regions, the
density of the gas is typically high, and therefore gas can
radiatively cool and supply fuel for star-formation within the galaxy.

Several methods can be used to probe the hot gas surrounding galaxies,
such as observations of the thermal Sunyaev-Zel'dovich (SZ) effect
\citep{sunyaev:1970, vanderlinde:2010, planck:2013, anderson:2015} and
X-ray emission, on which we focus in this work. In the analytic model
of \cite{white:1991} the typical temperature of the gaseous atmosphere
is $T > 10^{6} \units{K}$ in a halo of mass, $10^{12}
\units{\msun}$. Therefore, the atmosphere radiates as it cools through
line-emission and continuum, with significant emission in the soft
X-ray energy band, $0.5-2.0 \units{keV}$ \citep[see][for a recent
review]{putman:2012}.

The first attempts to detect soft X-ray emission from hot, gaseous,
coronae around nearby, late-type galaxies were made with the
\textit{ROSAT} X-ray satellite, but no convincing evidence for it was
found \citep{benson:2000}. These observations instead provided upper
limits for the X-ray luminosity of the coronae, which were almost two
orders of magnitude lower than the analytical predictions of
\cite{white:1991}. The origin of the overestimate can be traced back
to the assumption in that paper that the gas has an isothermal density
profile whereas, as found by \cite{crain:2010} in the GIMIC
cosmological hydrodynamics simulations, the gas is more diffuse due to
the removal of low entropy gas by star formation and, most importantly
on galactic scales, energy injection from supernovae

Advances in X-ray detector sensitivity in the \textit{XMM-Newton} and
\textit{Chandra} telescopes led to the first detections of diffuse
X-ray emission around nearby, late-type galaxies
\citep{strickland:2004, wang:2005, tullman:2006, li:2007, owen:2009,
  sun:2009}. While detections of diffuse, X-ray coronae are now
commonplace, a variety of studies have found that they appear as thick
discs \citep{strickland:2004}, trace galactic outflows of H$\alpha$
\citep{tullman:2006}, and have total luminosities that correlate
strongly, and positively, with the recent star-formation rate
\citep{li:2013}. Highly star-forming galaxies, such as M82, also
exhibit filamentary X-ray structures above, and below, the galactic
plane \citep{strickland:2004, li:2012}. These observations and
inferred correlations suggest that the dominant source of the X-ray
emission around local disc galaxies is gas heated by supernovae (SNe)
feedback\footnote{We follow the incorrect, but now common usage of the
  word ``feedback'' to refer to the energy emitted by supernovae or by
  AGN.}, rather than gas cooling from an accreted quasi-hydrostatic
atmosphere.
This interpretation, however, is in conflict with more recent deep
\textit{XMM-Newton} observations of NGC 6753 \citep{bogdan:2017},
NGC 1961 \citep{anderson:2016} and NGC 891
\citep{hodges-kluck:2018}. These data provide compelling
evidence for the existence of hot, low-metallicity atmospheres of gas
that are consistent with accretion from the IGM and subsequent
shock-heating to the virial temperature of the halo. Nevertheless, the source
of X-ray emission around late-type galaxies, like the Milky Way (MW)
and M31, remains controversial. 

A further important unknown is the mass fraction in hot
atmospheres. In galaxy clusters, the halos are almost `baryonically
closed' \citep{white:1993,vikhlinin:2006, pratt:2009, lin:2012}, such
that their baryon-to-dark-matter mass ratios within the virial radius
are approximately equal to the mean cosmic ratio,
$f_{b} = \rho_{b}/\rho_{m}$, where $\rho_{b}$ and $\rho_{m}$ are the
baryonic and total matter density of the universe, respectively. This
ratio is taken to be $0.157$ \citep{planck:2013}. However,
the baryon fractions of halos around $L^\ast$ galaxies, appear to be much
lower than this and thus a significant fraction of the baryons appear to be
`missing' from the haloes \citep{bregman:2018}. In this paper we also
study the gas fraction of halos, $f_{\mathrm{gas}}$, which we define
as the ratio of the mass of hot gas to halo mass, normalised by the
mean cosmic baryon fraction, $f_{b}$. Galaxy halos also contain
cold and warm, which is detectable in absorption studies of galaxies
\citep{tumlinson:2017}, however in the halo mass range we focus on, hot
gas dominates the total baryonic mass.

In this paper, we use the large volume, cosmological hydrodynamical
simulation suite EAGLE \citep{schaye:2015, crain:2015, mcalpine:2016}
to probe the origin, mass and X-ray properties of the hot gaseous
atmospheres surrounding present-day disc galaxies. We use the
simulations to investigate the relative contributions of the accreted 
shocked-heated gas and winds heated by feedback to the X-ray
luminosity, \lx, of hot gas atmospheres. We then compare the soft
X-ray scaling relations \lxmvir\ and \lxmstar\ for a large sample of
simulated disc galaxies with observational data over a wide range of
halo masses, $(10^{11}-10^{15}) \units{\msun}$. We further examine the
effect of the varying gas fraction of halos on the slope and
the normalisation of the \lxmvir\ relation.

The EAGLE simulations have previously been shown to reproduce a wide
range of observations of real galaxies, such as low-redshift hydrogen
abundances \citep{lagos:2015, bahe:2016}, evolution of galaxy stellar
masses \citep{furlong:2015} and sizes \citep{furlong:2017},
star-formation rates and colours \citep{trayford:2015, trayford:2017}
along with black hole masses and AGN luminosities
\citep{guevara:2016}.

The paper is structured as follows. In Section~\ref{sec:methods} we
discuss the simulations, the selection of our simulated disc galaxy
sample, the method used to calculate the X-ray luminosities and
present analytic predictions for the \lxmvir\ relation. In
Section~\ref{sec:baryon_census} we perform a baryon census of the
EAGLE reference simulation. We then investigate the origin of the
X-ray emission in Section~\ref{sec:xray_origin} and attempt to
understand how this depends on both the spatial region around the
galaxy and the halo mass. In Section~\ref{sec:xray_observations} we
compare the results of the simulations to a range of observational
data. We further investigate the effects of AGN feedback on the X-ray
and gas properties of halos by comparing simulations with differing
AGN models. In Section~\ref{sec:xray_scaling} we investigate the
\lxmvir\ scaling relations in the EAGLE simulations and compare them
to our analytical predictions. In Section~\ref{sec:baryon_from_xray} we discuss how
to infer the gas fractions of halos from the measured \lxmvir\
relation. Section~\ref{sec:missing_feedback} we introduce the
``missing feedback'' problem and use high cadence simulation outputs
to identify the physical origin in the simulations, before concluding
in Section \ref{sec:conclusions}.

\section{Methods and background}\label{sec:methods}
\subsection{Numerical simulations}\label{sec:sims}

We make use of the large volume cosmological hydrodynamical
simulations, EAGLE \citep[Evolution and Assembly of GaLaxies and their
Environments,][]{schaye:2015, crain:2015, mcalpine:2016}, to follow
the evolution of galaxies and their gaseous atmospheres. The EAGLE
simulations adopt a $\Lambda$CDM cosmology with the parameters of
\cite{planck:2013} listed in Table~1 of \cite{schaye:2015}.

The EAGLE simulations were performed with a highly modified version of
the \textsc{GADGET-3} \citep{springel:2005}. The fluid properties are
evolved using the particle-based smoothed particle hydrodynamics (SPH)
method \citep{lucy:1977, gingold:1977}. The EAGLE simulations use a
pressure-entropy formulation of SPH \citep{hopkins:2013}, with
artificial viscosity and conduction switches \citep{price:2008,
  cullen:2010} which, when combined, are referred to as
\textsc{ANARCHY}.

The EAGLE simulations include a variety of sub-resolution baryonic
physics relevant to galaxy formation such as radiative gas cooling
\citep{wiersma:2009}, star formation \citep{schaye:2008}, metal
enrichment \citep{wiersma:2009b}, black-hole seeding, active galactic
nuclei (AGN) feedback \citep{springel:2005, guevara:2015} and feedback
from stellar evolution \citep{vecchia:2012}. The subgrid physics model
has several parameters which were tuned to reproduce the present-day
stellar mass function, the size distribution of disc galaxies and the
relationship between galaxy stellar mass and central black hole
mass. It is important to note that the gas properties of the
simulations were not tuned to match any observations and, as a result
they are genuine predictions of the galaxy formation model.

The prescription for energy injection from SNe feedback is critically
important as SNe can deposit large amounts of thermal and kinetic
energy into the gas immediately surrounding the galaxy. Observations
show that the energy feedback from SNe can heat the gas to
temperatures, $T \geq 10^{7} \units{K}$, which is hot enough to
contribute to the X-ray luminosity within the galactic halo
\citep{strickland:2007}. Observations suggest that this hot gas is
also able to drive winds via conversion of thermal to kinetic energy,
which can propagate to large radii enriching and heating material
\citep{rupke:2018}. It is not currently possible to resolve individual
stars or SNe within large volume cosmological simulations; instead, a
single stellar particle of mass, $\approx 10^{6} \units{\msun}$,
represents a population of stars. The simulations then require a
prescription for energy deposition and metal enrichment from each star
particle onto the surrounding gas which is tuned in order to reproduce a
variety of observed galaxy properties.

Traditionally, hydrodynamical simulation codes have injected the
energy from SNe events within a single stellar population (SSP),
represented by a star particle, over a large mass of gas
\citep{schaye:2008, creasey:2011, keller:2014}. For a standard stellar
initial mass function (IMF) there is $\approx 1$ supernova per
$100 \units{\msun}$ of initial stellar mass. Assuming the energy from
SNe within this SSP is injected into a mass of gas equal to the
initial stellar mass formed leads to high initial temperatures for the
gas, $T \geq 10^{7} \units{K}$. In the case where the same amount of
energy is distributed over a much larger mass of gas, the temperature
increase experienced by the gas will be much lower, which reduces the
cooling time. If the cooling time is significantly shorter than the
sound-crossing time of the gas, the energy injection is no longer able
to drive winds efficiently. \cite{schaye:2008} showed that in this
case, simulations are unable to reproduce observed star-formation
rates and stellar masses of galaxies. There are several different SNe
energy injection techniques that have been used to address this 
`over-cooling' problem, including injecting energy in kinetic form,
depositing the energy thermally while disabling radiative cooling
for a short period and dumping the energy both thermally and
kinetically \citep[e.g.][]{navarro:1993, hernquist:2003, schaye:2008}.

The EAGLE galaxy formation model injects the energy from SNe entirely
thermally \citep{schaye:2015}. However, instead of distributing the
energy evenly over all of the neighbouring gas particles, it is
injected into a small number of neighbours \textit{stochastically}
\citep{vecchia:2012}. This method allows the energy per unit mass,
which corresponds to the temperature change of a gas particle, to be
defined. In the simulations, each gas particle heated by SNe feedback
is always subject to the same temperature increase, namely
$\Delta T_{\rm{SN}} = 10^{7.5} \units{K}$.

The AGN feedback proceeds in a very similar manner to the SNe
feedback.  In the EAGLE simulations black hole (BH) seeds are placed
at the centre of halos with a mass greater than
$6.7 \times 10^{9} \units{\msun}$ that do not already contain a
BH. The rate of gas accretion by BHs is modelled using the local gas
density, velocity and angular momentum, along with the mass of the
BH. As the BH accretes gas, it accumulates a reservoir of energy equal
to the energy of the gas mass accreted multiplied by the radiative
efficiency, which is taken to be $10\%$. When the BH has stored
sufficient energy, it can stochastically increase the temperature of
some of the neighbouring gas particles by a temperature of
$\Delta T_{\rm{AGN}}$. In the reference simulations, particles are
subject to a temperature change of
$\Delta T_{\rm{AGN}} = 10^{8.5} \units{K}$; however this can be
varied. A lower temperature change means particles are heated more
often, a (higher) temperature change, less often. As previously
mentioned, the parameters of the AGN and SNe feedback are calibrated
so as to reproduce the galaxy stellar-to-halo mass relation.

In this work, we use several of the EAGLE simulations, described in
Table~\ref{tab:eagle_runs}. We focus on the largest volume simulation,
Ref-L100N1504, which uses the EAGLE reference subgrid physics
model. This simulation is of a periodic cube of side length
$100 \units{cMpc}$, populated with $N=1504^3$ collisionless dark
matter particles and an equal number of baryonic particles. The impact
of AGN feedback is investigated by using two variants of the reference
simulation. In the NoAGN simulation the AGN feedback has been
disabled, whereas, in the AGNdT9 simulation, the AGN feedback has been
modified such that each feedback event leads to a temperature change
of $\Delta T_{\rm{AGN}} = 10^{9} \units{K}$. The remainder of the
physical parameters, including mass resolution, remain the
same. Further details of these different EAGLE simulations may be
found in \cite{crain:2015}.

\subsection{Halo and galaxy identification}\label{sec:galaxy}

\begin{table}
    \begin{center}
        \begin{tabular}{ l | c c c c }
            \hline
            Name & N & Mass & Box Size & $\Delta$ T$_{\rm{AGN}}$ \\
            units & & $\units{\msun}$ & $\units{cMpc}$ & $\units{K}$ \\
            \hline
            Ref-L100N1504 & $1504^3$ & $1.81 \times 10^6$ & $100$ & $10^{8.5}$ \\
            NoAGN-L050N0752 & $752^3$ & $1.81 \times 10^6$ & $50$ & - \\
            AGNdT9-L050N0752 & $752^3$ & $1.81 \times 10^6$ & $50$ & $10^{9.0}$ \\
        \end{tabular}
    \end{center}
    \caption{Parameters of the EAGLE simulations analysed in this work. The columns are the name of the simulation, the number of dark matter particles (which is initially equal to the number of gas particles), initial gas particle mass, the length of the side of the box, and the temperature change induced by AGN feedback, if AGN feedback is enabled. The runs are named such that the prefix, e.g., `Ref` refers to the subgrid physics parameters followed by \textit{LXXXNYYYY} where \textit{XXX} is the side-length of the cube in $\units{Mpc}$ and \textit{YYYY}$^{3}$ is the number of dark matter particles.}  
    \label{tab:eagle_runs}
\end{table}

In this section we describe the procedure to identify galaxies in the
EAGLE simulations at redshift, $z=0$. We also describe the
morphological criteria that we employ to select only galaxies with
significant disc components.

Dark matter halos are identified using a Friends-of-Friends (FoF)
algorithm with a linking length of $0.2$ times the mean dark matter
interparticle separation \citep{davis:1985}.  The gas, stars and BHs
are associated with the FoF group of their nearest dark matter
particle if it belongs to a FoF group. The constituent self-bound
substructures (subhalos) within a FoF group are identified using the
\textsc{SUBFIND} algorithm applied to both dark matter and baryonic
particles \citep{springel:2001, dolag:2009}.

In this work we focus on \textit{centrals}, which are the central
galaxies in a dark matter halo. These are identified as the most
massive individual subhalos with a centre of mass lying within a $20$
proper kpc (pkpc) radius of the centre-of-mass of the host FoF
group. If no such subhalo exists within the FoF group, we discard the
halo. We further require that galaxies be subhalos containing at least
$500$ star particles, in order to ensure we have a large sample of
star particles to use for morphological classification.

We compute the spherical overdensity mass \citep{lacey:1994} of each
FoF halo about the deepest particle within the potential of the
halo. We define the halo radius to be the spherical radius within
which the mean enclosed density is $\Delta$ times the critical density
of the universe, $\rho_c$. We generally adopt $\Delta = 200$ to define
virial quantities but we also use $\Delta = 500$ in some analyses to
allow a more appropriate comparison to the observational data of
\cite{anderson:2016}.

\subsection{Morphological and isolation selection}\label{sec:disc}

We characterise the morphology of galaxies by means of the
$\kappa_{\mathrm{rot}}$ parameter introduced by
\cite{sales:2012}. The parameter is defined as,
\begin{equation}
    \kappa_{\mathrm{rot}} = \frac{1}{K} \sum \frac{1}{2} m \Bigg  (\frac{j_{i, z}}{R_{i}} \Bigg)^2~,
    \label{eq:kappa}
\end{equation}
where $K$ is the total kinetic energy of the stellar particles, $m$ is
the mass of each stellar particle, $j_{i, z}$ is the $z$-component of
the specific angular momentum, R$_{i}$ is the 2D projected radius from
the $z$-axis and the sum is performed over all stellar particles
within the galaxy. The galaxy is oriented such that the total angular
momentum of all stellar particles within the galaxy lies along the
$z$-axis. We consider all stellar particles within a spherical radius
of $30 \units{pkpc}$ around the most bound stellar particle to be
associated with the galaxy. In general
$\kappa_{\mathrm{rot}} \approx 1$ for discs with perfect circular
motions, whereas $\kappa_{\mathrm{rot}} \approx 1/3$ for non-rotating
systems. A visual inspection of the stellar projections, both edge-
and face-on, of galaxies in \cite{sales:2012} suggests that
$\kappa_{\mathrm{rot}} \geq 0.50$ corresponds to galaxies that exhibit
clear disc morphology.

We also apply an isolation criterion to the sample of disc galaxies
analysed. We only select galaxies in halos which do not intersect a
sphere of radius, $3 R_{200}$, spanned by any of their neighbouring
halos. These galaxies are undesirable as their X-ray emission is often
dominated by the hot gas associated with their (more) massive
neighbours.

We can increase the stringency of the disc criterion by increasing the
required threshold value of $\kappa_{\mathrm{rot}}$, but in this work,
we define disc galaxies to be those with
$\kappa_{\mathrm{rot}} \geq 0.50$; this results in a sample of
$\approx 5000$ disc galaxies in the fiducial simulation,
Ref-L100N1504. Increasing $\kappa_{\mathrm{rot}}$ further reduces the
size of our sample significantly, but does not change our main
results; decreasing the value leads to the selection of a large number
of galaxies that have no observable disc component in projection,
e.g. elliptical and irregular galaxies which would not be appropriate
for comparison with most of the observational samples considered
here.

Galaxies with halo mass, $\geq 10^{12.5} \units{\msun}$, are not
subject to any morphological criterion. This is because very few
halos of this mass host disc galaxies in EAGLE and the primary
observational data to which we compare  in this mass range makes
no selection for disc galaxies.

\subsection{Computing the X-ray emission}\label{sec:xray_calc}

The X-ray emission of galactic gas coronae is calculated in
post-processing. The X-ray luminosity of each gas particle is
calculated independently using the precomputed lookup tables from the
Astrophysical Plasma Emission Code \textsc{APEC 3.0.1} data
\citep{smith:2001, foster:2012}. The data assumes that the gas is an
optically thin plasma in collisional ionisation equilibrium. The total
cooling rate is computed for individual elements as a function of 
photon energy. The total cooling rate per element is computed by
integrating over a given range of photon energies,
$0.5-2.0\units{keV}$. We then calculate the total cooling rate by
summing the overall contribution of each element,

\begin{equation}
    \Lambda_{\mathrm{X}}=\sum_{i} X_i \Lambda_{\mathrm{X},i}~,
\label{eq:lambda_xray}
\end{equation}
where $X_i$ is the ratio of the element abundance in the gas relative
to the solar abundance, $\Lambda_{\mathrm{X}, i}$ is the cooling rate
of the gas in the X-ray band, $0.5-2.0 \units{keV}$ (hereafter, soft
X-ray), assuming solar abundances, and $\Lambda_{\mathrm{X}}$ is the
total soft X-ray luminosity. We use the solar abundances of
\cite{anders:1989} to normalise the abundances in our simulations.

The summation in Eqn.~\ref{eq:lambda_xray} is performed over nine
elements: hydrogen, helium, carbon, nitrogen, oxygen, neon, magnesium,
silicon and iron, which are independently tracked within the
simulation. The total X-ray luminosity from the hot halo is the sum of
the X-ray emission of all the particles within a given spatial
region. This X-ray calculation does not include contributions from
non-gaseous X-ray sources within galaxies, e.g. X-ray binaries. The
effect of these sources is discussed when comparing to suitable
datasets.

We calculate the projected coronal soft X-ray luminosity around a galaxy in the
following way. We select all gas particles within a sphere of radius,
$R_{200}$, centred on the centre-of-mass of the halo. The total X-ray
emission is then calculated by summing the X-ray emission from all the
gas particles within a 2D annulus through the sphere.

The APEC software, used to generate the X-ray luminosity emission tables, was not used to generate the emission tables used for calculating cooling rates within the simulations; instead, the simulations used tables generated by CLOUDY. This use of two different emission tables can lead to an error in the computations of X-ray luminosities. For example, if the APEC emission tables predict higher emissivity than CLOUDY, then the X-ray luminosity will be overpredicted since a self-consistent simulation using the APEC cooling tables would have less gas at the same density and temperature due to the faster cooling. This effect is not important for gas with long cooling times, but it may be significant for gas with shorter cooling times. We expect this to be a small effect in this work as the X-ray emission is dominated by hotter, slower cooling gas $\approx 10^{6} \units{K}$.

\subsection{\texorpdfstring{\lxmvir\ relation}%
                               {X-ray-to-halo mass relation}}\label{sec:lx_mhalo_relation}

 For gas in hydrostatic equilibrium in a
 dark matter halo, $L_{\mathrm{X}} \propto M^\alpha_{\mathrm{vir}}$. If the
 bolometric X-ray emission is dominated by thermal bremsstrahlung radiation, halos
 have a constant gas fraction, and the gas density profiles are self-similar, then
 the slope of the scaling relation has the classical value, $\alpha = 4/3$
 \citep{kaiser:1986,sarazin:1986}. However, in the halo mass and energy range considered in this work, the first two of these assumptions are not valid.

 We can derive the scaling relation in the halo mass range, $10^{12.0}-10^{13.5}
 \units{\msun}$, and  energy band, $0.5-2.0 \units{keV}$, following the work of
 \cite{bohringer:2012}:
\begin{equation}
    L_{\mathrm{X}} \propto f_{\mathrm{gas}}^2 \Lambda(T_{\mathrm{vir}}) M_{\mathrm{vir}}~,
    \label{eq:lx_halo_mass_approx}
\end{equation}
where $f_{\mathrm{gas}}$ is the gas fraction of the halo normalised by
the cosmic baryon to total mass ratio; $\Lambda(T_{\mathrm{vir}})$ is the
cooling function of the gas as a function of the virial temperature,
$T_{\mathrm{vir}}$; and $M_{\mathrm{vir}}$ is the virial mass
\footnote{We use the term ``virial mass'' to refer to both $M_{200}$
  and $M_{500}$ but the distinction should be clear in the appropriate
  context. These quantities scale proportionally and agree to
  within $10\%$.} In the halo mass range,
$10^{11.5}-10^{13.5} \units{\msun}$, the cooling rate,
$\Lambda_{\mathrm{X}}$, is approximately proportional to the
temperature of the gas, as demonstrated in
Appendix~\ref{appendix:xray}. As the virial temperature scales as 
$T_{\mathrm{vir}} \propto M_{\mathrm{vir}}^{2/3}$, then
$\Lambda \propto M_{\mathrm{vir}}^{2/3}$. In the case when the baryon
fraction can be expressed as a power law,
$f_{\mathrm{gas}} \propto M_{\mathrm{vir}}^{\beta}$,
Eqn.~\ref{eq:lx_halo_mass_approx} simplifies to,
\begin{equation}
    \begin{split}
    L_{\mathrm{X}} &\propto M^\alpha_{\mathrm{vir}}\\
    &\propto f_{\mathrm{gas}}^2  M_{\mathrm{vir}}^{5/3}  \\
    &\propto M_{\mathrm{vir}}^{2 \beta} M_{\mathrm{vir}}^{5/3} \propto M_{\mathrm{vir}}^{5/3 + 2 \beta}~.\\
    \end{split}
    \label{eq:lx_halo_mass_power}
\end{equation}
This derivation assumes that the halo gas is at the virial temperature
of the halo and that the gas density profiles are self-similar as a
function of halo mass. We explore the validity of these assumptions in
Appendix~\ref{appendix:self-similar}. It is often common to consider 
the X-ray luminosity as a function of stellar mass,
e.g. $L_{\mathrm{X}} \propto M^{\alpha^*}_{\mathrm{star}}$. 

When the gas fraction is constant as a function of halo mass, the
slope of the \lxmvir\ relation is $\alpha = 5/3$. The increased
steepness, compared to the classical self-similar prediction,
$\alpha = 4/3$, is due to the scaling of the cooling function in this
halo mass range and energy band considered. It is also clear from
Eqn.~\ref{eq:lx_halo_mass_power} that an increase in the halo gas
fraction, $f_{\mathrm{gas}}$, with increasing halo mass, will result
in a steeper slope for the \lxmvir\ relation.

\section{Baryon census}\label{sec:baryon_census}

\begin{figure}
    \includegraphics[width=1.00\linewidth]{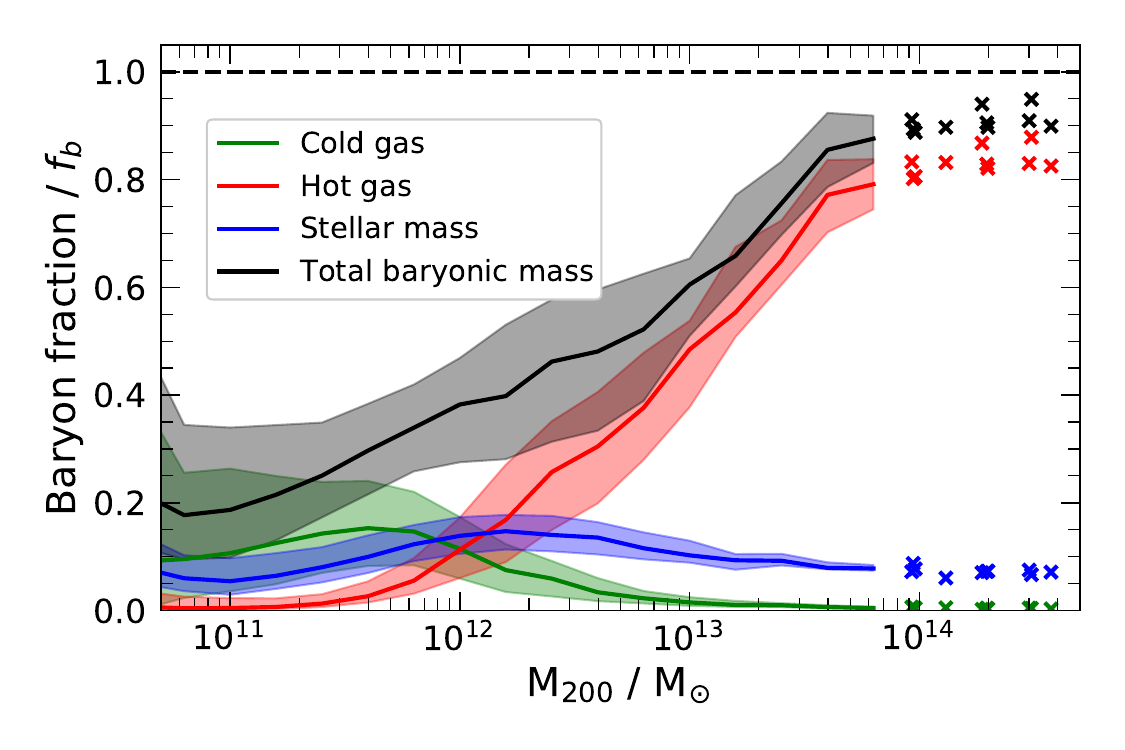}
    \caption{The baryon fraction of a sample of disc galaxies,
      selected from the EAGLE $(100 \protect\units{Mpc})^3$ reference
      model, Ref-L100N1504, as a function of the halo mass at redshift
      $z=0$. The values are normalised to the mean baryon fraction of
      the universe, $f_{b}$. The lines show the median, in mass bins
      of $0.20 \units{dex}$, of the baryon fraction of stars, cold gas
      ($T < 5 \times 10^{5} \units{K}$), hot gas
      ($T > 5 \times 10^{5} \units{K}$) and all baryonic particles in
      green, blue, red and black respectively. In bins with less than
      five objects, we show the results as individual crosses.  The
      shaded bands enclose the $15$th and $85$th percentiles.}
    \label{fig:eagle_fb}
\end{figure}

The baryon fraction of our sample of disc galaxies is shown in
Fig.~\ref{fig:eagle_fb} as a function of halo mass. Here, we plot the
baryon fraction of stars, cold gas
($T \leq 5 \times 10^{5} \units{K}$), hot gas
($T > 5 \times 10^{5} \units{K}$) and total baryons. The distinction
between hot and cold gas is motivated in Appendix~\ref{appendix:xray},
where we show the X-ray cooling function as a function of gas
temperature. In general, gas below a temperature of
$5 \times 10^{5} \units{K}$ has negligible X-ray emission in the
energy band range, $0.5-2.0 \units{keV}$, on which we focus in this
work.

Fig.~\ref{fig:eagle_fb} demonstrates that the baryon content of
low-mass halos, $\mvir < 10^{12} \units{\msun}$, is dominated by stars
and cold gas. The total baryon fraction within these halos is much
lower than the mean cosmic baryon-to-dark matter ratio. This low
baryon fraction is the consequence of the efficient feedback which
heats gas and can eject it to distances well beyond the virial radius
of the halo \citep{schaller:2015, mitchell:2019}. In low mass halos we see a negligible contribution of mass from hot gas. This is not surprising since even if these halos hosted gaseous halos of accreted gas, their typical temperature  would be $\leq 5 \times 10^5 \units{K}$, which we classify as cold gas.
Since our sample of galaxies was selected
to be isolated, dynamical interactions, such as stripping, should not
affect the baryon content of the halos.

As the halo mass increases above a critical mass,
$\mvir \approx 10^{12} \units{\msun}$, we find a rapid increase in the
contribution of hot gas and in the total baryon fraction. 
The increase in the amount of hot gas is due to the virial temperature
of the halos increasing to a value that exceeds the threshold for our definition of hot gas, $5 \times 10^{5} \units{K}$. The increase in
total baryon fraction is likely due to the deepening of
the gravitational potential well of the halo, which increases its
ability to retain gas heated by feedback. As the halo mass increases further, to
$\approx 3 \times 10^{12} \units{\msun}$ and above, the hot gas
becomes the predominant mass component within the halo. The cooling
time of the accreted gas is now so long that, after shock-heating, the
gas forms a hot, quasi-hydrostatic atmosphere at (approximately) the
virial temperature of the halo \citep{larson:1974,
  white:1978,white:1991}.

In the EAGLE simulations, the feedback efficiencies of SN and AGN,
which regulate the stellar mass and halo baryon fractions, cannot be
predicted from first principles. As discussed in
Section~\ref{sec:methods}, adjustable parameters are calibrated to
match observed present-day galaxy properties, such as the galaxy
stellar mass function. The simulations also broadly reproduce the
stellar-to-halo mass relationship inferred from abundance matching
\citep{behroozi:2013, moster:2013}. However, EAGLE slightly
underpredicts the stellar mass at low halo mass and slightly
overpredicts it at high stellar mass \citep[see Fig.~8
of][]{schaye:2015}. In contrast to the galaxy stellar mass function,
the baryon fraction of halos are direct, non-calibrated predictions
of the subgrid physics model. 

Unfortunately, the gas fractions in real halos, at a given halo or
stellar mass, are uncertain. Previous studies of hot gas in the Milky
Way have suggested that the mass of hot gas within the virial radius
ranges between $(2 - 13) \times 10^{10} \units{\msun}$
\citep{nicastro:2016}, with various other estimates falling within
this large range \citep{gupta:2012, faerman:2017, bregman:2018}. These
constraints suggest that hot gas can account for a fraction between
$(10-100)\%$ of the baryon budget of the MW. In higher mass objects,
e.g. clusters, the baryon fractions are better constrained and the hot
gas makes up between $(70-100)\%$ of the total baryon budget
\citep{vikhlinin:2006, pratt:2009, lin:2012}. The results of
Fig.~\ref{fig:eagle_fb} are consistent with current observational
constraints.

\begin{figure*}
    \includegraphics[width=0.99\textwidth]{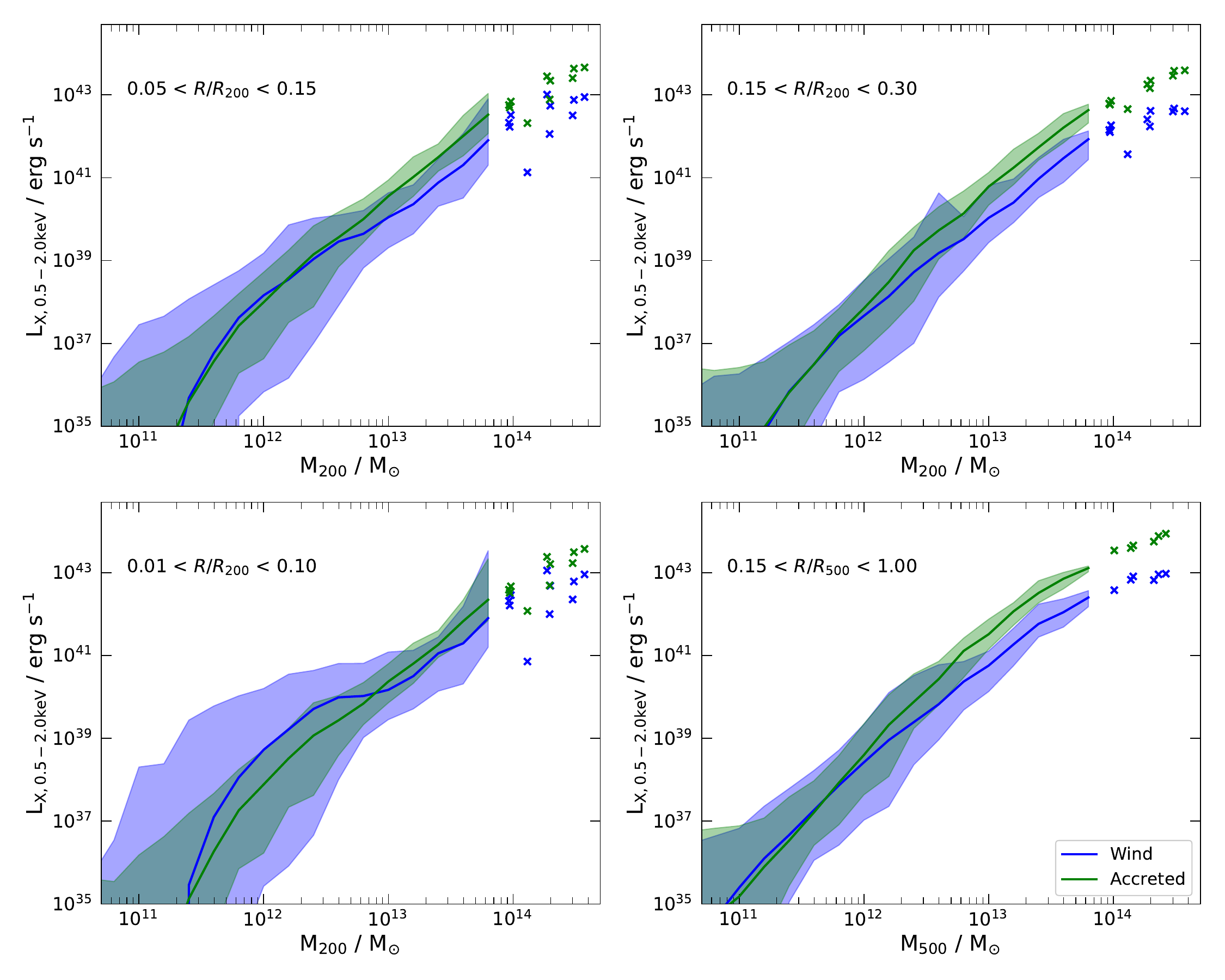}
    \caption{The soft X-ray luminosity,
      $0.5$-$2.0 \protect\units{keV}$, within different annuli around
      the galactic centre at $z=0$ for a sample simulated disc galaxies
      in the EAGLE simulation, Ref-L100N1504, as a function of the
      halo mass. The top left, top right, bottom left and bottom right
      show the X-ray luminosity in the annuli
      $0.05 < r/R_{200} < 0.15$, $0.15 < r/R_{200} < 0.30$,
      $0.01 < r/R_{200} < 0.10$ and $0.15 < r/R_{500} < 1.00$,
      respectively. In all panels the simulated median X-ray
      luminosity is calculated for both accreted gas (green) and wind
      (blue) in halo mass bins of $0.20 \units{dex}$ and shown with
      the solid line. The luminosities of halos in bins sampled by
      fewer than five galaxies are shown individually. The shaded
      bands enclose the $15$th and $85$th percentiles within the
      same mass bins.
      The halo mass in each panel is taken to be $M_{500}$, except for the
      lower-left panel where we use $M_{200}$, to provide the best
      comparison to the respective observations in Section~\ref{sec:xray}.
      }
    \label{fig:eagle_scaling_prist_wind}
\end{figure*}

\section{The X-ray luminosity}\label{sec:xray}

Fig.~\ref{fig:eagle_fb} shows that galaxies in halos of mass
$\geq 10^{12} \units{\msun}$ in the EAGLE simulations are surrounded
by hot gaseous coronae. While these gaseous atmospheres make up the
majority of the hot gas mass in the halo, it is not clear whether they
are the primary source of X-ray emission in these halos. Winds driven
by feedback may dominate the X-ray emission as they are typically very
hot, dense and metal-rich. Therefore, we must first identify the
origin of the dominant X-ray emitting gas before meaningful comparisons
can be made between simulated and observed X-ray halo luminosities.

\subsection{The origin of the X-ray emission}\label{sec:xray_origin}

We analyse the contribution of wind and accreted gas to the total
X-ray emission by considering the history of every gas particle within
the virial radius of the halos.
This gas can be classified into two categories: interstellar medium
(ISM) and circumgalactic medium (CGM). The ISM is typically the
high-density star-forming gas within the galaxy, whereas the CGM is
the surrounding halo gas. In the EAGLE simulations, the ISM is usually
defined as gas with a physical atomic number density,
$n_{H} > 0.1 \units{cm^{-3}}$, while all other gas within the halo is
considered to be part of the CGM.

The typical definition of ``wind'' is gas that has been ejected from
the ISM into the CGM or beyond. We, therefore, distinguish ``wind'' from
``accreted'' gas particles according to their ISM and halo accretion
histories.  Specifically, we compare the time since the particle was
last classified as ISM, $t_{\mathrm{ISM}}$, to the time since the
particle was most recently accreted into any FoF halo,
$t_{\mathrm{accretion}}$. We calculate $t_{\mathrm{accretion}}$ using
high-cadence ($240$ equally-spaced outputs from when the age of the
universe is $1 \units{Gyr}$ to $13.85 \units{Gyr}$) simulation outputs
to calculate the time when each gas particle was most recently {\em
  not} associated with a FoF group, $t_{\mathrm{accretion}}$. The
$t_{\mathrm{ISM}}$ is tracked by the simulation code which stores the
most recent time, if ever, when a gas particle was at an atomic number
density higher than the threshold for star formation,
$n_{\mathrm{H}} \approx 0.1 \units{cm^{-3}}$.

The case $t_{\mathrm{ISM}} > t_{\mathrm{accretion}}$ indicates
that a gas particle was accreted by the present-day halo
\textit{after} it was last in the ISM of a galaxy.
This suggests the gas particle was within the ISM of a galaxy at an
earlier time and was then ejected from that halo before joining the
IGM of a progenitor of the present-day halo. This gas, in the context
of the present-day host halo, represents \textit{accretion}. By
contrast, $t_{\mathrm{ISM}} < t_{\mathrm{accretion}}$ indicates the
gas particle has been in the ISM of the galaxy since it was last
accreted into a progenitor of the present-day halo. This gas was
accreted, cooled and joined the ISM before being ejected, probably as
a result of feedback, into the CGM. We, therefore, classify these
particles as \textit{wind}.

In Fig.~\ref{fig:eagle_scaling_prist_wind} we plot the coronal soft
X-ray luminosity within several 2D annuli for a sample of disc
galaxies selected from the Ref-L100N1504 simulation, as a function of
the halo mass. These annuli are chosen as they allow a direct
comparison with observations which we present in
Section~\ref{sec:xray_observations}. We show the X-ray luminosity
contribution from gas particles classified as \textit{accretion} and
\textit{wind} separately, in green and blue, respectively.

The largest contribution from wind is seen in the bottom-left panel of
Fig.~\ref{fig:eagle_scaling_prist_wind} which shows the central region
of the halo. In this annulus, the X-ray emission from wind can be up
to two orders of magnitude more luminous than emission from accreted
gas, with the median X-ray emission from wind about an order of
magnitude more luminous than from accreted gas for halos of mass
$\leq 10^{13} \units{\msun}$. However, above this mass, the median
X-ray emission from accreted gas is typically more luminous, but there
is still significant scatter likely reflecting different recent
star-formation rates. The \textit{wind} contributes disproportionately
to the X-ray luminosity compared to its contribution to the gas mass,
which is not shown. This is because the gas defined as \textit{wind}
is consistently hotter, more metal-rich and denser than accreted gas
\citep{crain:2010}. The upper percentile of the X-ray emission from
\textit{wind} in the bottom-left panel of
Fig.~\ref{fig:eagle_scaling_prist_wind} shows that the
$L_{\mathrm{X}}-M_{\mathrm{halo}}$ relationship flattens at low halo
masses. This reflects the increase in X-ray luminosity in these halos,
$M_{\mathrm{vir}} \leq 10^{13} \units{\msun}$, whereas there is no
increase in the median, total X-ray luminosity of higher mass halos,
$M_{\mathrm{vir}} \geq 10^{13} \units{\msun}$, due to \textit{wind} in
any of the annuli considered. In the lowest mass halos, $\mvir \leq 10^{12} \units{\msun}$, the $L_{\mathrm{X}}-M_{\mathrm{halo}}$ relation steepens again as below this halo mass there is very little accreted gas sufficiently hot to produce soft X-rays. Therefore, the X-ray emission drops rapidly with decreasing halo mass for low mass halos. In these same halos the recent star formation rates are not converged in the reference simulation, increasing (decreasing) the resolution of the simulation increases (decreases) the star formation rate. As the the X-ray luminosity in these halos is dominated by wind (see Fig.\ref{fig:eagle_scaling_prist_wind}), this means the X-ray luminosity may not be converged in these halos. Therefore, the X-ray luminosity predictions for these low mass halos should be used with caution. In more massive halos, $\mvir \geq 10\times10^{12} \units{\msun}$ both the recent star formation rate, and X-ray luminosity, are well converged with varied resolution.

In radial regions further out, we see a reduction in the contribution
of X-ray luminosity from \textit{wind} at a given halo mass. This
reduction happens because feedback processes, which generate winds,
are concentrated in the central regions of the halo. In the upper- and
lower-right panels of Fig.~\ref{fig:eagle_scaling_prist_wind}, where
the inner region is excised, we see that the X-ray emission from
\textit{wind} in lower mass halos,
$M_{\mathrm{vir}} \leq 10^{12} \units{\msun}$, is of the same order as
the total X-ray emission within the halo. However, for higher mass
halos, the median X-ray emission from accreted material is
significantly more luminous than \textit{wind}. As the halo mass
increases further, the fraction of the X-ray emission produced by
accreted gas converges to $\approx 100 \%$ at a halo mass of
$\approx 10^{13} \units{\msun}$. These outer annuli are therefore
ideal for probing quasi-hydrostatic, accreted halos without pollution
from the X-ray luminous, metal-rich wind. It should be noted, however,
that the X-ray surface brightness is much lower in the outer regions,
and thus difficult to observe around individual
galaxies. Nevertheless, \cite{oppenheimer:2020} argue that 4-year
eROSITA observations should be able to detect X-ray emission out to
$\approx 200 \units{kpc}$ for stacked data around halos of mass,
$\geq 10^{12} \units{\msun}$.

We note that the X-ray emission from diffuse gas which we classify as
\textit{wind} in the EAGLE simulations may not be representative of
the X-ray emission around real highly star-forming galaxies. This
X-ray emission is the result of a subgrid feedback model which injects
thermal energy directly into gas particles. The direct heating of gas
particles within the ISM of a galaxy, by both AGN and SNe feedback,
leads to star particles of mass, $M \approx 10^{6} \units{\msun}$,
with high-metallicities, $Z \approx \units{Z_{\odot}}$, high densities
$n_{\mathrm{H}} \geq 0.1 \units{cm^{-3}}$ and temperatures exceeding
$10^{7} \units{K}$. The feedback model is, of course, just an
approximation and its realism can only be established by comparison
with observations, such as those in
Section~\ref{sec:xray_observations} below.

Interactions between wind and accreted gas complicate the identification
of the origin of X-ray emission. In halos of mass $\sim 10^{12} \units{\msun}$ we expect that the accreted gas is shock-heated to the virial temperature \cite{white:1991}. However, in some cases, there could be additional shocks caused by wind-halo interaction, whereby hot outflows heat the gas in the CGM. As a result energy injected by supernovae may be emitted by particles which we have classified as accreted. This can lead to an overestimation of the energy emitted by accreted gas. This effect will be largest in lower mass halos for two reasons. The first is that the velocity at which the ejected gas encounters the infalling gas is larger in small mass haloes. Secondly, in small halos, the X-ray emission produced by wind and accreted gas appear to be comparable, whereas in massive halos, accreted gas dominates the total X-ray emission.


In summary, we find that a large fraction of the X-ray emission in the
central region, $R < 0.10 R_{\mathrm{vir}}$, of halos of
$M_{\mathrm{vir}} \leq 10^{13} \units{\msun}$ is produced by
\textit{wind}, which is the direct result of feedback processes
associated with stellar evolution and AGN.  However, when the central
region is excised, the gas classified as \textit{accretion} becomes
the predominant source of X-ray emission in halos of mass
$\geq 10^{12} \units{\msun}$. This shows that accreted X-ray emitting
coronae do exist around halos of mass $M_{\mathrm{vir}} \geq 10^{12}$,
as predicted by \cite{white:1991}, at least within the EAGLE
hydrodynamical simulations.

\subsection{The X-ray scaling relations}\label{sec:xray_scaling}

\begin{figure*}
    \includegraphics[width=1.00\textwidth]{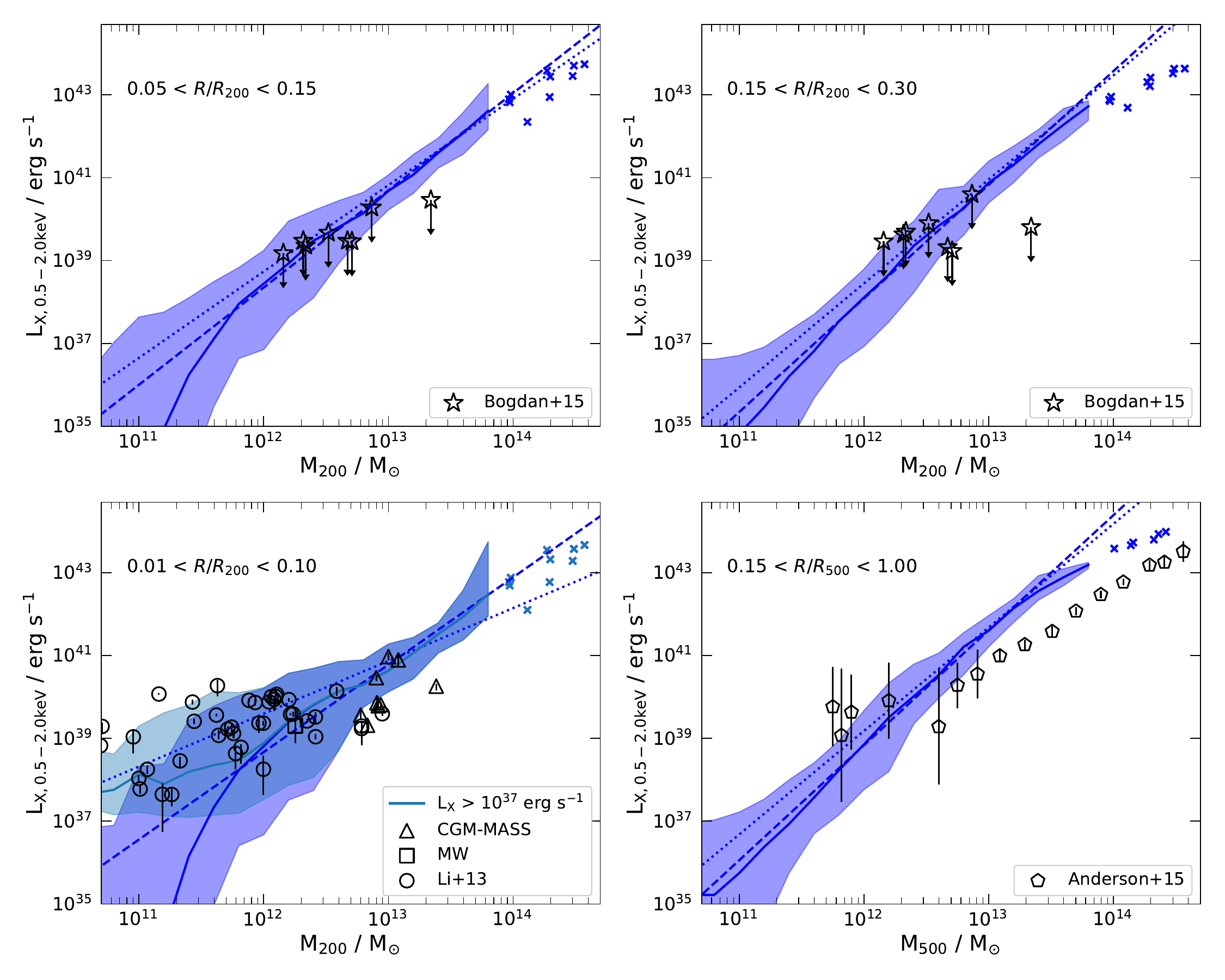}
    \caption{The soft X-ray luminosity,
      $0.5$-$2.0 \protect\units{keV}$, within different annuli around
      the galactic centre at $z=0$ for a sample of simulated disc
      galaxies in the EAGLE simulation, Ref-L100N1504, as a function
      of halo mass. The top-left, top-right, bottom-left and
      bottom-right panels show the X-ray luminosity in the annuli
      $0.05 < r/R_{200} < 0.15$, $0.15 < r/R_{200} < 0.30$,
      $0.01 < r/R_{200} < 0.10$ and $0.15 < r/R_{500} < 1.00$,
      respectively. These regions are chosen to allow appropriate
      comparisons with the observational results of
      \protect\cite{bogdan:2015} (top panels), \protect\cite{li:2017}
      (bottom left) and \protect\cite{anderson:2015} (bottom right)
      which are shown as the black data points with errorbars. In all
      panels the simulated median X-ray luminosity is calculated in
      halo mass bins of $0.20 \units{dex}$. The luminosities of halos
      in bins sampled by fewer than five galaxies are shown
      individually. The shaded bands enclose the $15$th and $85$th
      percentiles within the same mass bins. The dashed (dotted) lines show the
      best-fit to median (mean) X-ray luminosity in the mass range
      $10^{11.5} < M_{\mathrm{vir}} / \units{\msun} < 10^{13.5}$. The
      teal region in the lower-left panel shows the same sample with
      all galaxies of X-ray luminosity below $10^{37} \units{erg s^{-1}}$ excluded.
      The halo mass in each panel is taken to be $M_{500}$, except for the
      lower-left panel where we use $M_{200}$, to provide the best
      comparison to the respective observations.
      }
    \label{fig:eagle_scaling}
\end{figure*}

\begin{figure*}
    \includegraphics[width=1.00\textwidth]{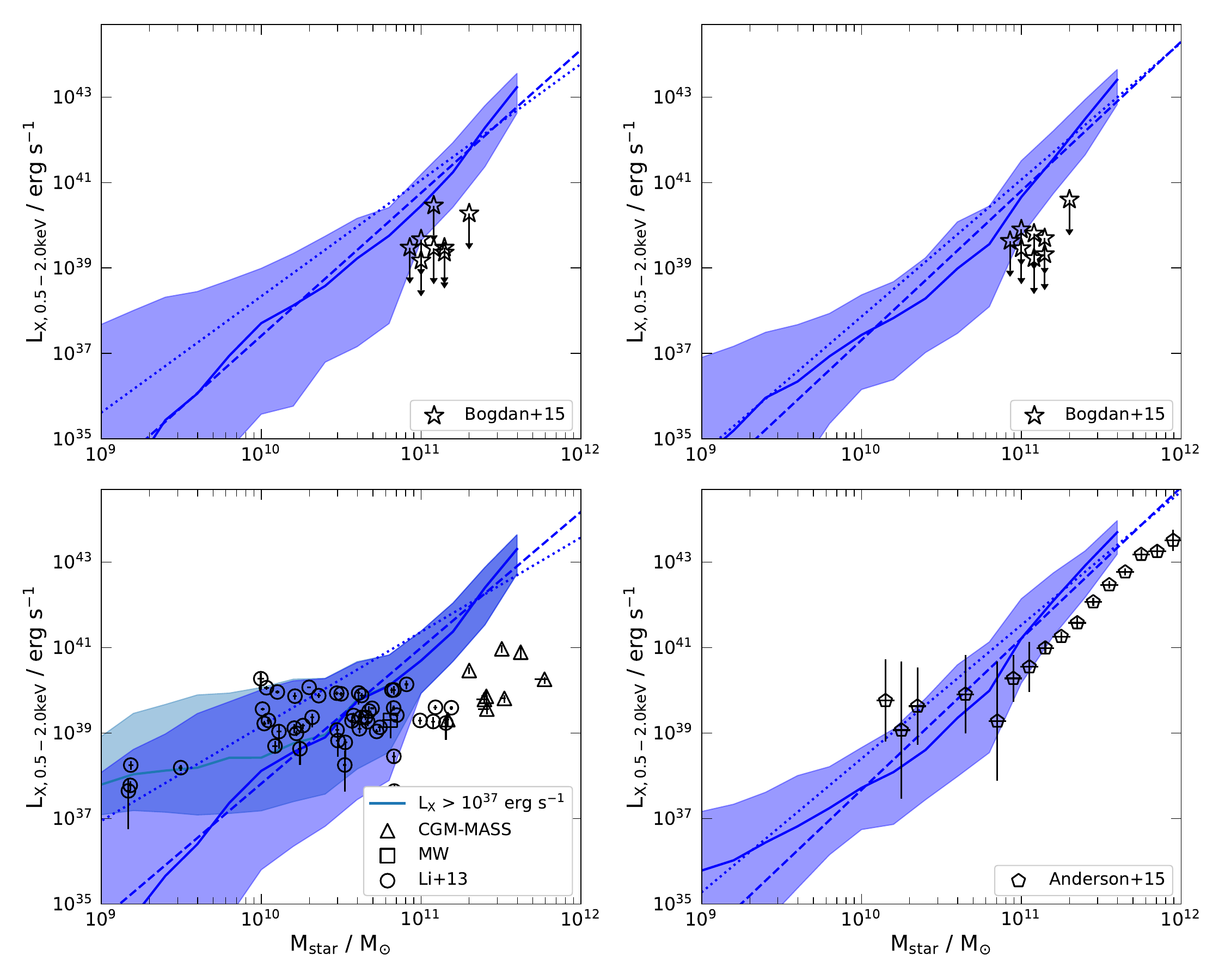}
    \caption{The same as Fig.~\protect\ref{fig:eagle_scaling}, but
      with the X-ray luminosity plotted as a function of the {\em
        stellar} mass of the central galaxy. The stellar mass is
      defined to be the total mass of stars within a 3D sphere of
      radius $30~\units{pkpc}$. The dashed (dotted) lines show the
      best-fit to median (mean) X-ray luminosity in the mass range
      $10^{9.5} < M_{\mathrm{star}} / \units{\msun} < 10^{11.5}$. The
      teal region in the lower-left panel shows the sample with all
      galaxies of X-ray luminosities below
      $10^{37} \units{erg s^{-1}}$ excluded.}
    \label{fig:eagle_scaling_stellar}
\end{figure*}

The coronal soft X-ray luminosity in 2D annuli for our sample of disc
galaxies selected from the reference EAGLE simulation is shown in
Figs.~\ref{fig:eagle_scaling} and \ref{fig:eagle_scaling_stellar} as a
function of halo and stellar mass respectively. These annuli are the same
as those considered in Fig.~\ref{fig:eagle_scaling_prist_wind}. In
Figs.~\ref{fig:eagle_scaling} and \ref{fig:eagle_scaling_stellar} we
also plot observational data from \cite{bogdan:2015},
\cite{li:2017} (bottom-left panel) and \protect\cite{anderson:2015}
(bottom-right panel).

The general trend in Figs.~\ref{fig:eagle_scaling} and
\ref{fig:eagle_scaling_stellar} is that the X-ray luminosity increases
with both halo and stellar mass. We can also see that the scatter in
X-ray luminosity, at either a fixed halo or stellar mass, increases
with decreasing mass. The larger scatter at low mass may be
interpreted as the result of the greater importance of
non-gravitationally heated gas, which is concentrated around the
central regions and not directly related to the halo mass. As the halo
mass increases, the scatter in the X-ray luminosity decreases, 
reflecting the increasing importance of gravitational heating on the
X-ray luminosity.

As we can see in Fig.~\ref{fig:eagle_scaling} the \lxmvir\
relationship is fairly well described by a single power law over
approximately three orders of magnitude in halo mass. We plot the
best-fit line to the relation in all panels of the figure using linear
regression on the logarithm of the median of the X-ray luminosity as a
function of the logarithm of the median halo mass. We find the
exponent of the scaling relation, $L_X \propto M^\alpha_{\rm vir}$, 
for the median X-ray luminosity to be $\alpha = 2.2 \pm 0.1, 2.6 \pm 0.1, 2.1 
\pm 0.1, 2.8 \pm 0.1$ in the halo mass range, $(10^{11.5}-10^{13.5}) 
\units{\msun}$, for the upper-left, upper-right, lower-left and lower-right panels,
respectively. Repeating this process with a best fit to the mean
X-ray luminosity gives $\alpha = 1.8 \pm 0.1, 2.4 \pm 0.1, 1.3 
\pm 0.1, 2.5 \pm 0.1$ in the same mass range for the upper-left,
upper-right, lower-left and lower-right panels, respectively.
We apply the same methodology to the \lxmstar\ relation,
shown in Fig.~\ref{fig:eagle_scaling_stellar}, in the stellar-mass
range $(10^{9}-10^{11}) \units{\msun}$ and find the exponent of the
relation to be,
$\alpha^* = 3.4 \pm 0.1, 3.5 \pm 0.2, 3.2 \pm 0.1, 3.5 \pm 0.2$ for the
median X-ray luminosity in the upper-left, upper-right, lower-left
and lower-right panels, respectively. These slopes reduce to
$\alpha^* = 2.7 \pm 0.2, 3.2 \pm 0.2, 2.2 \pm 0.2, 3.2 \pm 0.2$ in
the same mass range for the mean X-ray luminosity, respectively.

The \lxmvir\ and \lxmstar\ scaling relations are systemically flatter
in the two left panels of Figs.~\ref{fig:eagle_scaling} and
\ref{fig:eagle_scaling_stellar} which probe the inner region of the
halo. We also see that the difference in the slope of the scaling
relations, \lxmvir\ and \lxmstar\, between the best fit to the median
and the best fit to the mean X-ray luminosity is much larger in these two panels.

The origin of the flatter slope, and the discrepancy between the
mean and median X-ray luminosity over a small range of halo mass, is due to the enhanced X-ray emission
in low mass halos, $\sim (10^{11}-10^{13}) \units{\msun}$ in the inner
region compared to the outer region. The increased X-ray luminosity
within the central regions is caused by feedback, as shown in
Fig.~\ref{fig:eagle_scaling_prist_wind}. However, not all low-mass
galaxies have recent star formation. Thus, we see a scatter of up to four orders of magnitude in the X-ray luminosity in these halos.
Therefore, a small sample of highly star-forming, X-ray luminous galaxies
are able significantly to increase the mean X-ray luminosity, while having
a smaller impact on the median X-ray luminosity, at a fixed halo mass. As
the median X-ray luminosity is less affected by feedback from recent star formation,
we focus on the median X-ray luminosity in the remainder of this paper.
In the outer regions, displayed in the two right panels of
Fig.~\ref{fig:eagle_scaling}, we find that the slope of the scaling relation, \lxmvir, is steeper than the analytical prediction for self-similar
gaseous halos, $\alpha = 4/3$, presented by \cite{kaiser:1986} and
\cite{sarazin:1986} and the $1.8$ value inferred from the observations
of \cite{anderson:2015}. We investigate the origin of the steeper
slope of the \lxmvir\ scaling relation in
Section~\ref{sec:effects_of_agn}.

\subsection{Comparison to observations}\label{sec:xray_observations}

\begin{figure*}
    \includegraphics[width=1.00\textwidth]{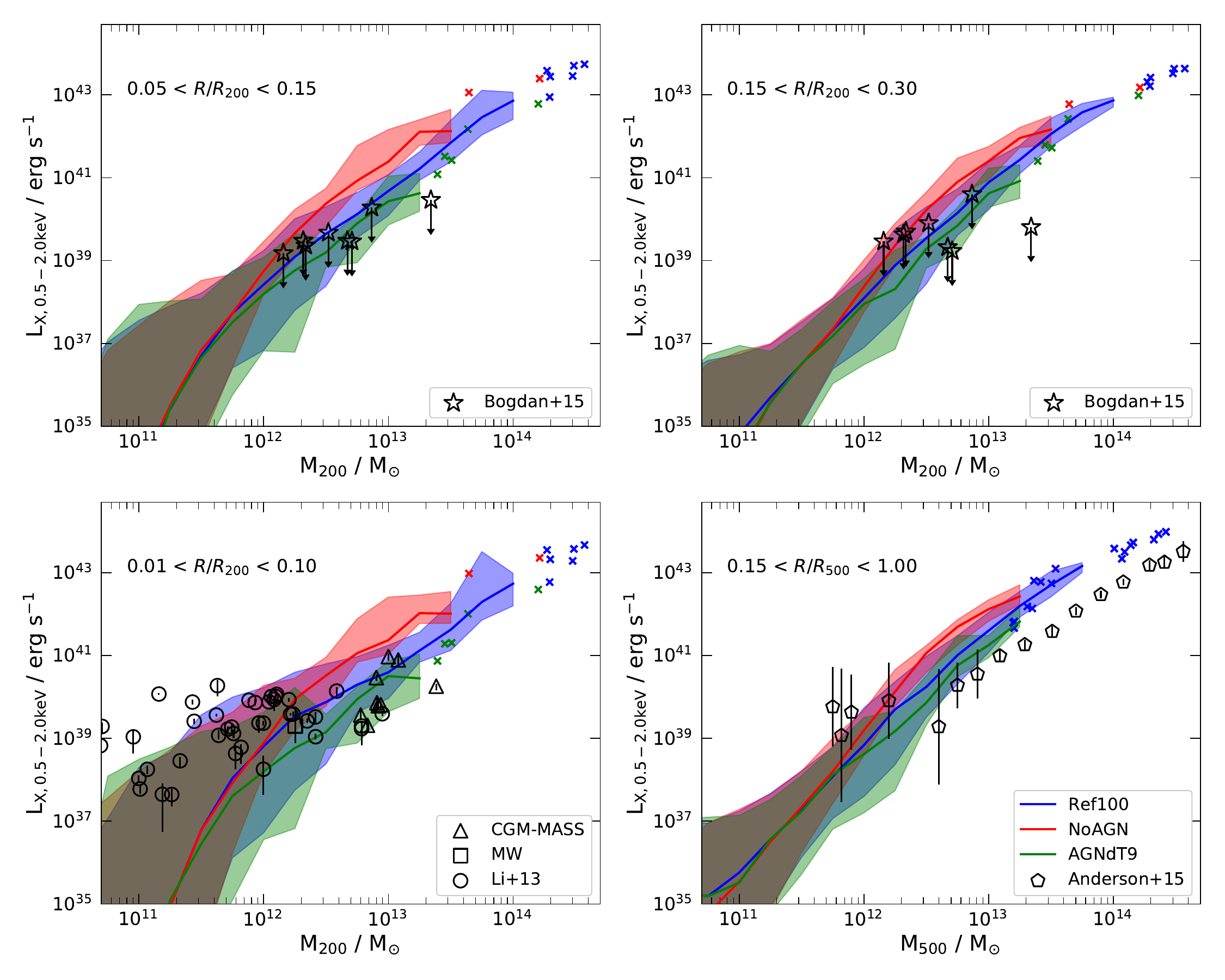}
    \caption{The same as Fig.~\protect\ref{fig:eagle_scaling}, but
      showing the X-ray luminosity for three different samples of discs
      galaxies taken from the EAGLE simulations Ref-L100N1034 (blue),
      NoAGN-L050N0752 (red), AGNdT9-L050N0752 (green).  This
      comparison shows the impact of varying the AGN feedback model on
      the X-ray luminosity to halo mass relationship.}
    \label{fig:eagle_scaling_agn}
\end{figure*}

We now calculate the coronal X-ray luminosity from the simulated
galaxies in a way that allows a fair comparison to observations, that
is, within the same annulus and energy range. However, there are still
limitations in the direct comparison of simulated and observed X-ray
luminosities. For example, when calculating the X-ray emission in an
annulus, we only include gas within a sphere of the virial
radius. However, in real observations, the line-of-sight X-ray
emission may be contaminated by non-gaseous X-ray sources such as
X-ray binaries in the galaxy, or unrelated background and foreground
objects. A fraction of the coronal X-ray emission in the real universe
may be absorbed, particularly at lower energies. These effects are not
considered when calculating the X-ray emission from simulated
galaxies. As our analysis of the simulations is not
instrument-limited, we include galaxies with X-ray luminosities well
below the current detection threshold. The simulations thus contain
more low luminosity objects than observational samples, but we account
for that in the comparison with the data. 

In the top two panels of Fig.~\ref{fig:eagle_scaling} and
Fig.~\ref{fig:eagle_scaling_stellar} we compare the simulations to the
data presented by \cite{bogdan:2015}, who used \textsc{Chandra} to
search for soft X-ray emission around eight normal spiral galaxies.
No statistically significant diffuse soft X-ray emission was detected
around any of these galaxies. We therefore use the inferred $3\sigma$
upper limits for our comparison. As these observations excise the
central regions, any contamination unresolved X-ray point sources
should be small. 

The upper limits derived from the observations of \cite{bogdan:2015} generally overlap with the 15th to 85th percentiles predicted by EAGLE. However, these upper limits \cite{bogdan:2015} are consistently lower than the mean and median X-ray luminosity found in the simulations. Therefore, there does appear to be some evidence that the simulations are slightly, but significantly overpredicting the X-ray emission at large luminosities.
However, the X-ray luminosities as a function of stellar
mass in the simulations are significantly higher than the
observational limits, as may be seen in
Fig.~\ref{fig:eagle_scaling_stellar}. The reason for the discrepancy
between the results at fixed halo and at fixed stellar mass could be
due to two reasons. The first is that the stellar-to-halo mass
relationship in the EAGLE simulations could be incorrect. However,
abundance matching suggests that the EAGLE relation agrees well with
the data \citep{schaye:2015}. Secondly, it could be that the halo or
stellar masses inferred for the real galaxies is incorrect. The
stellar masses in \cite{bogdan:2015} are estimated from K-band
luminosities in the 2MASS survey, while the halo masses are estimated
from the circular velocity of gas in the disc, which is then converted
to a halo mass following \cite{navarro:1997}. In
Appendix~\ref{appendix:stellar_halo}, we compare the stellar-to-halo
mass relation of these observations with results from the EAGLE
simulation. In particular, the central panel of
Fig.~\ref{fig:stellar_halo_fixed} compares the stellar baryon fraction
as a function of halo mass in three different EAGLE simulations to the
observational estimates of \cite{bogdan:2015}. These observations imply
that the stellar mass accounts for between $15\%$ and $50\%$ of the
baryon budget of the halos. However, constraints from abundance
matching suggest that this value should be closer to $10\%$
\citep{moster:2013}. This is indicative of an overestimated stellar
mass, or an underestimated halo mass. As the stellar mass is more
directly inferred than the halo mass, we assume that it is the halo
masses that are underestimated. 

As the X-ray emission is calculated
in a halo mass dependent aperture, $(0.05 - 0.15)~R_{500}$, this can
lead to an incorrect value for the X-ray luminosity. Since the X-ray
emission is typically centrally peaked; an overestimated halo mass
implies that the central region excised would be too large, leading
to too low an X-ray luminosity. Thus, if the assumed halo masses are
indeed too large, the data points on the upper-left panel of
Fig.~\ref{fig:eagle_scaling} should be moved up and to the left,
reducing the discrepancy with the EAGLE simulations. The same data 
points would remain at the same stellar mass, but may also increase
in X-ray luminosity in Fig.~\ref{fig:eagle_scaling_stellar}.

Fig.~\ref{fig:eagle_scaling_prist_wind} shows that, at least within
the simulations, the observations of \cite{bogdan:2015} probe the
transition between \textit{wind} and \textit{accretion} dominated
X-ray emission. If the \cite{bogdan:2015} halo masses are underestimated,
their data will shift towards the \textit{accretion}-dominated
regime. The overestimated X-ray emission in the simulations could then
reflect excessive hot gas baryon fractions, or incorrect
thermodynamic properties for the gas.

In the bottom left panel of Figs.~\ref{fig:eagle_scaling} and
\ref{fig:eagle_scaling_stellar} we compare the predictions from the
EAGLE simulations to observations of massive, isolated spiral galaxies
homogeneously reanalysed by \cite{li:2017}. This sample includes the
detections of NGC 1961 \citep{bogdan:2013, anderson:2016} and NGC 6753
\citep{bogdan:2013}, which are referred to as ``massive spirals''. The
CGM-MASS sample of \cite{li:2017} and a measurement of the Milky Way
X-ray luminosity from \cite{snowden:1997} are also included alongside
the original sample of inclined disc galaxies presented by
\cite{li:2012} and \cite{li:2014}. These observations use
\textsc{Chandra} and \textsc{XMM-Newton} to probe the inner regions of
nearby halos.

We first consider the X-ray luminosity as a function of halo mass from
the innermost region. In this regime the observations are consistent,
in both the overall trend and scatter, with the simulated galaxies
within the halo mass range $(10^{11}-10^{13}) \units{\msun}$. The
agreement is particularly good when we exclude all simulated galaxies
with X-ray luminosities below $10^{37} \units{erg s^{-1}}$, as
shown by the teal region in Figs.~\ref{fig:eagle_scaling} and
\ref{fig:eagle_scaling_stellar}. This luminosity cut is consistent
with the observational limits of the data. For the highest mass halos
the simulations appear to overestimate the X-ray luminosity; however,
this is a tentative result given the small size of the observational
sample.  The simulations also reproduce the trend of the observed
\lxmstar\ relation in this innermost annulus.  In the lower-left panel
of Fig.~\ref{fig:eagle_scaling_stellar} we see that below a stellar
mass of $\approx 10^{11} \units{\msun}$ the simulations are consistent
with the observations, particularly once we exclude halos of
luminosity below $10^{37} \units{erg s^{-1}}$.

Fig.~\ref{fig:eagle_scaling_prist_wind} shows that the dominant source
of X-ray emission in the innermost region in the EAGLE simulations
is hot winds produced by feedback. We therefore suggest that the
observations of \cite{li:2017} are probing SNe-heated hot gas, rather
than the innermost region of a hot accreted halo.

Finally, the bottom right panel of Fig.~\ref{fig:eagle_scaling}
compares our simulations to the stacked X-ray observations of
\cite{anderson:2015}, which consist of a sample of approximately
$250000$ ``locally brightest galaxies'' from the Sloan Digital Sky
Survey. A more detailed description of the selection criteria is given
in \cite{planck:2013} but, in summary, galaxies are selected if they
are brighter than a threshold in extinction-corrected Petrosian
\textit{r}-magnitude band while also being the brightest object
within a $1 \units{Mpc}$ projected radius. These selection criteria
were chosen in an attempt to select a population of `central'
galaxies. Our sample of simulated galaxies is subject to a
conceptually similar selection process, in that we also choose
isolated galaxies. The X-ray emission from the real galaxies is
stacked in bins of stellar mass, and these stellar masses are
converted into halo masses. It should be noted that for this sample an
overdensity of $\Delta = 500$ is used to define the halo mass; to
facilitate a fair comparison we also compute this mass for our simulated
halos.

In the regime of MW-mass halos we find that the X-ray luminosities in
the EAGLE simulations are in good agreement with the observations at
both fixed halo and stellar mass. However, above this mass, the slopes
of the \lxmvir\ and \lxmstar\ relations in the simulations are steeper
than for the real galaxies. The overprediction of the X-ray luminosity
in the simulation peaks at a halo mass of
$\sim 3 \times 10^{13} \units{\msun}$ and then decreases to around a
factor of three for the most massive halos,
$\sim 10^{14} \units{\msun}$. \cite{wang:2016} recalibrated the
estimated halo masses of the \cite{anderson:2015} sample using weak
lensing data. \cite{wang:2016} suggest that the halo masses from \cite{anderson:2015} are slightly too large, with an almost constant overestimation of between $0.05-0.10 \units{dex}$ (see Fig. 10, right panel of \cite{wang:2016}). Thus, using the \lxmvir\ slope of 1.9 for the \cite{anderson:2015} data we predict that correcting the halo masses would increase the X-ray luminosity, at a given halo mass, by a factor of approximately $\sim 1.2-1.5$. Furthermore, a decrease in the inferred halo mass would decrease the virial radius, and therefore decrease the size of the excised central region. This would increase the X-ray emission, as emission is centrally peaked. We expect that these effects may change the results by up to a factor of two when combined. Given the large dynamic range of the data, this correction does not significantly reduce the tension between the EAGLE simulations and observations.

\subsection{Effects of AGN}\label{sec:effects_of_agn}

In this section we use the three EAGLE simulations, Ref-L100N1034,
NoAGN-L050N0752 and AGNdT9-L050N0752, to investigate the effects of
varying the implementation of AGN feedback on the \lxmvir\
relationship. The three simulations were introduced in
Section~\ref{sec:methods}. We repeat the sample selection process
outlined in Section~\ref{sec:disc} independently in each
simulation. The X-ray luminosity as a function of halo mass in the
three simulations is displayed in Fig.~\ref{fig:eagle_scaling_agn} 
for the same spatial regions, and compared to the same observational
data as in Fig.~\ref{fig:eagle_scaling}.

In all four regions we see that the X-ray luminosity in lower mass
halos, $\leq 10^{12} \units{\msun}$, is unchanged by the variation of
the AGN feedback implementation. In the EAGLE galaxy formation model
AGN have little effect on galactic properties below this critical mass
\citep{schaye:2015,guevara:2016,bower:2017,davies:2019}. Observationally,
it is also known that galaxies of stellar mass below
$\sim 10^{10} \units{\msun}$, which corresponds to a halo mass of
$\sim 10^{12} \units{\msun}$, seldom host powerful AGN
\citep{kauffmann:2003}. Above this halo mass we see a general trend
across all the spatial annuli: halos with no-AGN feedback have higher
X-ray luminosities. The differences in the X-ray luminosity in
simulations with and without AGN peak in the
$(10^{12} - 10^{13}) \units{\msun}$ mass range. At higher masses,
$\sim 10^{14} \units{\msun}$, the results from all the simulations
appear to converge. However, this is a tentative result as the two
$(50 \units{Mpc})^3$ simulations have a small number of haloes in this
mass range.

The AGNdT9-L050N0752 simulation has a modified AGN feedback model in
which the change in temperature, $\Delta T_{\mathrm{AGN}}$, due to AGN
feedback is increased to $10^{9} \units{K}$. In this simulation we see
that the X-ray luminosity at fixed halo mass in the range,
$(10^{12} - 10^{13}) \units{\msun}$, is lower than in both the
reference and no-AGN models. The decrease in luminosity, which is
typically about $0.5 \units{dex}$, improves the agreement with the
observations in all spatial regions. This indicates that the AGN
feedback in the reference model is under efficient. \cite{schaye:2015}
also found that the modified AGN feedback in AGNdT9-L050N0752 improves
the agreement between simulated and observed X-ray emission for some
of the most massive objects, $M_{500} \geq 10^{13} \units{\msun}$.
Further to this, \cite{correa:2018} found an upturn in the
ratio of the cooling radius to the virial radius of high mass halos,
$\geq 10^{13} \units{\msun}$, within the EAGLE reference simulation.
This upturn has also been attributed to under efficient AGN feedback
in high-mass halos. The increase in the cooling radius can account
for the significant overprediction of the X-ray luminosity in the inner
regions of high-mass halos (as seen in the two left panels of
Fig.~\ref{fig:eagle_scaling}).

In the EAGLE model, the main effect of AGN feedback is to eject gas
beyond the virial radius of the halo, as we show in
Section~\ref{sec:baryon_from_xray}. This decreases the total hot gas
mass and thus gas density in the halo, thus decreasing the X-ray
luminosity at fixed halo mass. AGN feedback can also decrease the
SFR in the galaxy, which would reduce the X-ray emission from wind in
the innermost regions.
These results are consistent with those of \cite{bogdan:2015} who
analysed a sample of spiral galaxies in the \textsc{Illustris}
simulations. Their ``textbook'' \textsc{Illustris} spiral galaxies
undershoot the observed X-ray emission, a fact that \cite{bogdan:2015}
attributed to over-efficient radio-mode AGN feedback which acts to
reduce the baryon fraction.

It is also interesting to note that the FIRE simulations analysed in
\cite{voort:2016} do not include AGN feedback and recover the observed
X-ray emission in $(10^{12}-10^{13}) \units{\msun}$ more convincingly
than the EAGLE reference model. The reason for the improved agreement
in this case may the implementation of stellar feedback which, in the
FIRE simulations, can drive efficient winds, even in high mass
halos. It is clear from Fig.~2 in \cite{voort:2016} that the hot gas
baryon fraction within their halos is $\sim 0.25 f_{b}$ at a halo
mass of $\sim 10^{13} \units{\msun}$, which is lower than that
found in any of the EAGLE simulations we are considering.

\section{Estimating the gas fractions from the \texorpdfstring{\lxmvir}{X-ray-to-halo mass}~relation}\label{sec:baryon_from_xray}

\begin{table*}
    \centering
    \begin{tabular}{ l | c c c c }
        \hline
        Name & $L_{\mathrm{X}}-M_{500}$ & Predicted $f_{\mathrm{gas}}-M_{500}$ & Empirical $f_{\mathrm{gas}}-M_{500}$ \\
        & {$\alpha$} & {$\beta$} & {$\beta$} \\
        \hline
        Ref-L100N1504 & $2.68 \pm 0.10$ & $0.51 \pm 0.05$ & $0.58 \pm 0.02$ \\
        \hline
        NoAGN-L050N0752 &  $1.98 \pm 0.16$ & $0.16 \pm 0.08$ & $0.18 \pm 0.05$ \\
        \hline
        AGNdT9-L050N0752 &  $2.64 \pm 0.05$ & $0.48 \pm 0.03$ & $0.44 \pm 0.10$ \\
        \hline
    \end{tabular}
    \caption{Best-fit exponents of the $L_{\mathrm{X}}-M_{\mathrm{vir}}$ and $f_{\mathrm{gas}}-M_{\mathrm{vir}}$ relationships in the EAGLE simulations. The exponents are calculated in the halo mass range $10^{12.5}-10^{13.5}$ M$_{\odot}$.
      The columns give the name of the simulation, the exponent of the
      $L_{\mathrm{X}}-M_{\mathrm{vir}}$ relation, the predicted exponent of the $f_{\mathrm{gas}}-M_{\mathrm{vir}}$ relation and the exponent pf $f_{\mathrm{gas}}-M_{\mathrm{vir}}$ relation. The relations are fit within the halo mass range, $10^{12.5}-10^{13.5} \units{\msun}$.}
    \label{tab:eagle_fb_predictions}
\end{table*}

\begin{figure*}
    \includegraphics[width=1.00\textwidth]{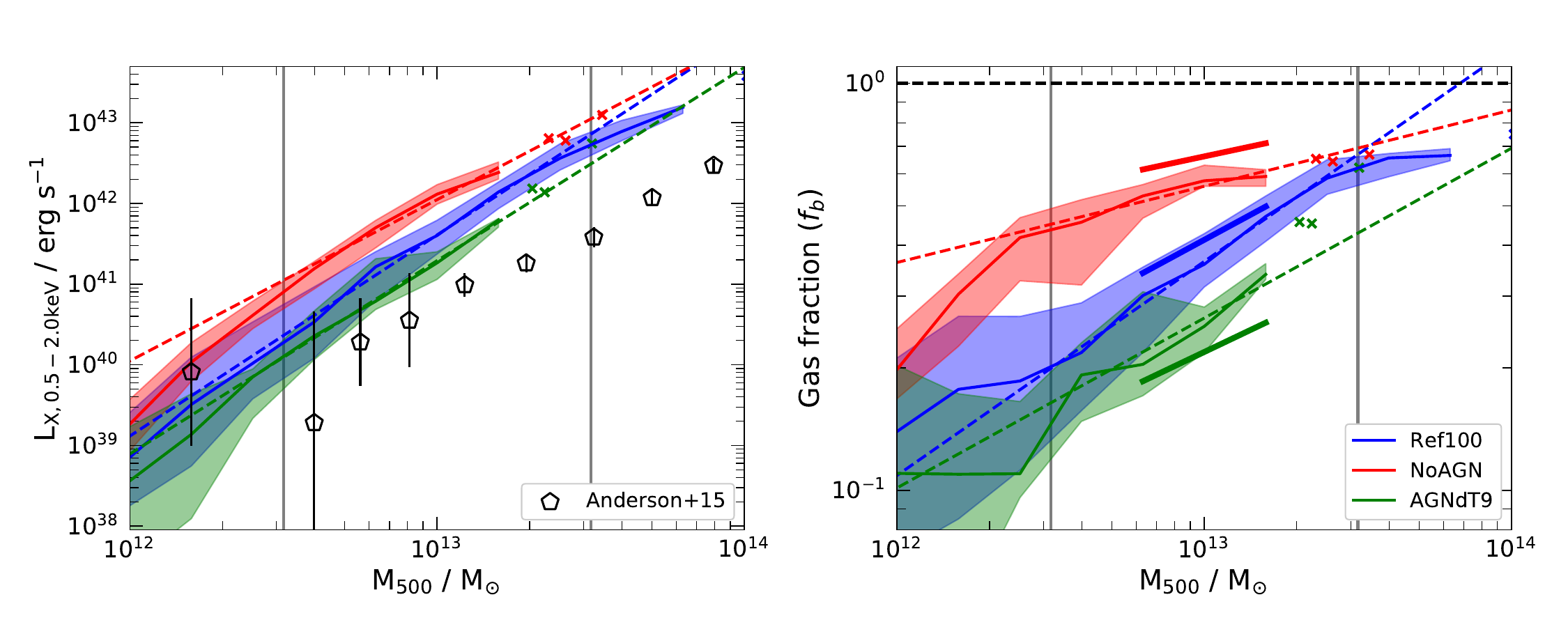}
    \caption{The coronal soft X-ray luminosity (left) and gas mass
      fraction (right) as a function of halo mass, $M_{500}$. In both
      panels we only consider gas within the 3D virial radius,
      $R_{500}$, and within the annulus $0.15 < R/R_{500} < 1.00$. The
      solid curves and shaded bands show the median and $15$th to
      $85$th percentiles in mass bins of $0.20 \protect\units{dex}$
      respectively. In bins that would enclose less than five objects
      we plot individual galaxies. The Ref-L100N1034, NoAGN-L050N0752
      and AGNdT9-L050N752 are shown by the green, blue and red curves
      respectively. We also add best-fit lines in both panels, which
      are shown with dashed lines. The model parameters of the best
      fits are shown in
      Table~\protect\ref{tab:eagle_fb_predictions}. The black vertical
      lines show the range of halo mass over which the best-fit is
      calculated. The thick solid line segments in the right panel show the 
      predicted slopes for the $f_{\mathrm{gas}}-M_{\mathrm{vir}}$ estimated
      from the slopes of the $L_{\mathrm{X}}-M_{500}$ relation in the left panel,
      following Eqn. \ref{eq:lx_halo_mass_power}.
    }
    \label{fig:eagle_fb_xray}
\end{figure*}

In Section~\ref{sec:lx_mhalo_relation} we showed that the dependence
of the baryon fraction on halo mass is encoded in the
$L_{\mathrm{X}}-M_{\mathrm{vir}}$ relation.  An increase in baryon fraction
with halo mass increases the slope of the relation and vice versa.  If
a power law can describe the gas fraction as a function of halo mass,
then Eqn.~\ref{eq:lx_halo_mass_power} can be used to calculate the
halo mass dependency of the gas fraction from the logarithmic slope of
the $L_{\mathrm{X}}-M_{\mathrm{vir}}$ relation. We evaluate this technique
using the X-ray luminosity, halo masses and gas fractions of our three
EAGLE simulations, Ref-L100N1034, NoAGN-L050N0752 and
AGNdT9-L050N0752. The variations in AGN feedback lead to noticeable
differences in baryon fraction and its dependence on halo mass for
halos of $M_{\mathrm{vir}}\geq 10^{12} \units{\msun}$.

In the left-hand panel of Fig.~\ref{fig:eagle_fb_xray} we plot the
coronal soft X-ray luminosity, in mass bins of $0.20 \units{dex}$,
within an annulus, $0.15 < R/R_{500} < 1.00$. The X-ray luminosity is
plotted as a function of halo mass, as in
Fig.~\ref{fig:eagle_scaling}, focusing on the mass range of interest,
$\mvir \geq 10^{12} \units{\msun}$, where significant gaseous halos
are present (see Fig.~\ref{fig:eagle_fb}). The right panel shows the
median gas fraction within the same annulus as a function of halo
mass, $M_{500}$.
As in Section~\ref{sec:xray_scaling}, we fit straight-lines to the
logarithm of both the median X-ray luminosity and the median gas
fraction plotted against the logarithm of the halo mass. We fit both
of these properties in the halo mass range,
$10^{12.5}-10^{13.5} \units{\msun}$, and tabulate the best-fit
parameters in Table~\ref{tab:eagle_fb_predictions}. The mass
range we fit in is different from that in Section~\ref{sec:xray_scaling}
and thus the slopes are slightly different.

We now use the slope of the $L_{\mathrm{X}}-M_{\mathrm{vir}}$ relation to
estimate the scaling of $f_{\mathrm{gas}}$ with $M_{\mathrm{vir}}$.
According to Eqn.~\ref{eq:lx_halo_mass_power},
$f_{\mathrm{gas}}\propto M^\beta_{\mathrm{vir}}$ with
$\beta = (\alpha_{\rm vir} - 5/3)/2$ (where $\alpha_{\rm vir}$ is
defined through $L_{\mathrm{X}}\propto M^{\alpha_{\rm
    vir}}_{\mathrm{vir}}$). We can calculate this scaling directly in
the simulations, as it is the slope of the best-fit line shown in the
right-hand panel of Fig.~\ref{fig:eagle_fb_xray}. We tabulate
$\alpha_{\rm vir}$, the predicted value of $\beta$ and the empirical
value of $\beta$ from the simulations in
Table~\ref{tab:eagle_fb_predictions}. The $1-\sigma$ errors of the
best-fit parameters, calculated from the covariance of the Jacobian,
are also included.

As Table~\ref{tab:eagle_fb_predictions} shows, the measured values of
the logarithmic slopes of the $f_{\mathrm{gas}}-M_{\mathrm{vir}}$
relation in all three simulations agree well with our predictions from
Section~\ref{sec:xray_scaling}, within $1.3 \sigma$.  The steepening of
the $L_{\mathrm{X}}-M_{\mathrm{vir}}$ above a slope of $5/3$ is accounted for
by the variation of the gas fraction as a function of halo mass. This
demonstrates that the X-ray emission is adequately described by
Eqn.~\ref{eq:lx_halo_mass_power} and that the variation of the gas
fraction makes an important and measurable contribution to the
logarithmic slope of the $L_{\mathrm{X}}-M_{\mathrm{vir}}$ relation. In
particular, we can distinguish between simulations with different AGN
implementations by the slope of the corresponding
$L_{\mathrm{X}}-M_{\mathrm{vir}}$ relation. In principle, this same methodology
can be applied to the real universe to understand how the gaseous
baryon fraction varies, from MW-mass halos to galaxy clusters.



The $L_{\mathrm{X}}-M_{\mathrm{vir}}$ relation flattens significantly
at large halo mass, $\geq 10^{13.5} \units{\msun}$. This is caused by
both a flattening of the soft X-ray cooling function with temperature
(see Appendix~\ref{appendix:xray}) and by the near-constant gas fraction
in high mass halos. These effects should combine to produce
an \lxmvir\ relation where $L_{\rm X} \propto M_{\rm vir}$ for halos of mass
$\gg 10^{13.5} \units{\msun}$ within the soft X-ray energy
band. However, since the simulations analysed here have no objects in
this mass range, we cannot test the validity of this prediction.

The data of \cite{anderson:2015} are consistent with a shallow
$L_{\mathrm{X}}-M_{\mathrm{vir}}$ scaling relation, with
$\alpha_{\rm vir} \approx 1.8$, in the halo mass range
$M_{500} \geq 10^{12} \units{\msun}$. \cite{anderson:2015} attributed
the increase from the logarithmic slope of $4/3$ predicted by
\cite{kaiser:1986} and \cite{sarazin:1986} to the effects of
non-gravitational heating from AGN. They suggested that
``self-regulated'' AGN feedback should increase the X-ray luminosity
of higher mass halos. In their picture, thermal instabilities due to
radiative cooling in the hot halo result in high BH accretion rates,
which cause energy build-up and subsequent feedback that heats the gas
in the central region. This process repeats cyclically. However, in
the EAGLE simulations, we find the opposite to be
true. Table~\ref{tab:eagle_fb_predictions} shows that the shallowest
slope of the \lxmvir\ relation occurs in the NoAGN-L050N0752
simulation.  AGN feedback in the other two simulations significantly
increases the logarithmic slope of the \lxmvir\ relation. We suggest
that the primary effect of AGN feedback is to \textit{decrease} the
X-ray emission, at fixed halo mass, particularly in lower mass halos
of $M_{200} \approx (10^{12}-10^{13}) \units{\msun}$. This is due to
the ejective nature of AGN feedback which reduces both the mass and
the density of the X-ray emitting gas.

A reinterpretation of the data of \cite{anderson:2015}, shown in the left
panel of Fig.~\ref{fig:eagle_fb_xray}, in the context of
Eqn.~\ref{eq:lx_halo_mass_power} suggests that the halo gas fraction
is approximately independent of halo mass, $\beta \approx 0.1$. When
comparing the simulation results with measurements derived from
stacked observational data, it is more appropriate to compare to the
mean of the simulated data rather than the median. We showed in
Section~\ref{sec:xray_scaling} that the slope of the \lxmvir\ relation
decreases by approximately $0.3$ when fit to the mean, rather than the
median, X-ray luminosity in the annulus $0.15 < R/R_{500} < 1.00$.
This suggests that the low value of $\beta$, which we inferred from
the \cite{anderson:2015} results could be higher, $\beta \approx 0.25$.
This result is more consistent with the gas fraction variation in the
NoAGN EAGLE simulation which predicts the MW hot halo hosts approximately
$(30-40)\%$ of the total halo baryon budget.

In summary, in the EAGLE simulations, we find that the steepening of
the \lxmvir\ relation above $5/3$ is due to the variation of the halo
gas fraction with halo mass. Table~\ref{tab:eagle_fb_predictions}
demonstrates that the slope of the $f_{\mathrm{gas}}-M_{\mathrm{vir}}$
relation can be robustly and precisely extracted from observations of
the \lxmvir\ relation in the halo mass range
$10^{12}-10^{13.5} \units{\msun}$. This same methodology can be
applied to the real universe to constrain the gas fraction of
halos. As the $f_{\mathrm{gas}}-M_{\mathrm{vir}}$ relation is strongly
affected by AGN feedback, these constraints will provide insight on
the scale and extent of AGN driven winds.

\begin{figure}
    \includegraphics[width=1.00\linewidth]{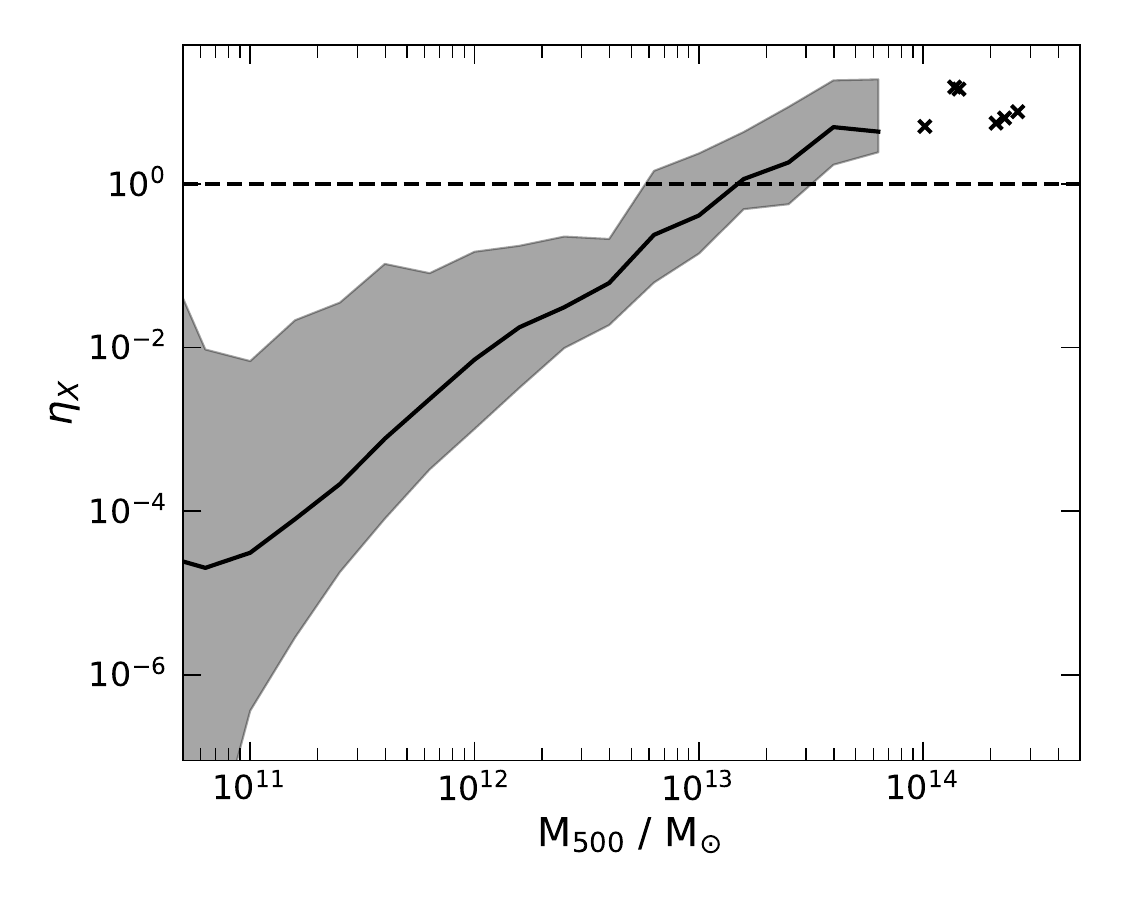}
    \caption{The X-ray coupling efficiency, $\eta_{X}$, as a function
      of halo mass in the EAGLE reference simulation for a sample of
      isolated disc galaxies. The X-ray coupling efficiency is defined
      as the ratio of the observed X-ray luminosity to the rate of
      energy input due to SNe, $L_{\mathrm{X}} / \dot{E}_{\mathrm{SN}}$. The
      power injected by SNe, $\dot{E}_{\mathrm{SN}}$, can be calculated
      in the simulations from the mean star formation rate in the last
      $250 \units{Myr}$ and the mean energy injected per unit of
      initial stellar mass formed. The black line shows the median
      X-ray coupling efficiency, in halo mass bins of
      $0.20 \units{dex}$, where the X-ray luminosity has been
      calculated from all the gas within the virial radius. The shaded
      bands enclose the $15$th and $85$th percentiles.}
    \label{fig:eta_m200}
\end{figure}

\section{Missing feedback problem}\label{sec:missing_feedback}

\begin{figure*}
    \includegraphics[width=1.00\textwidth]{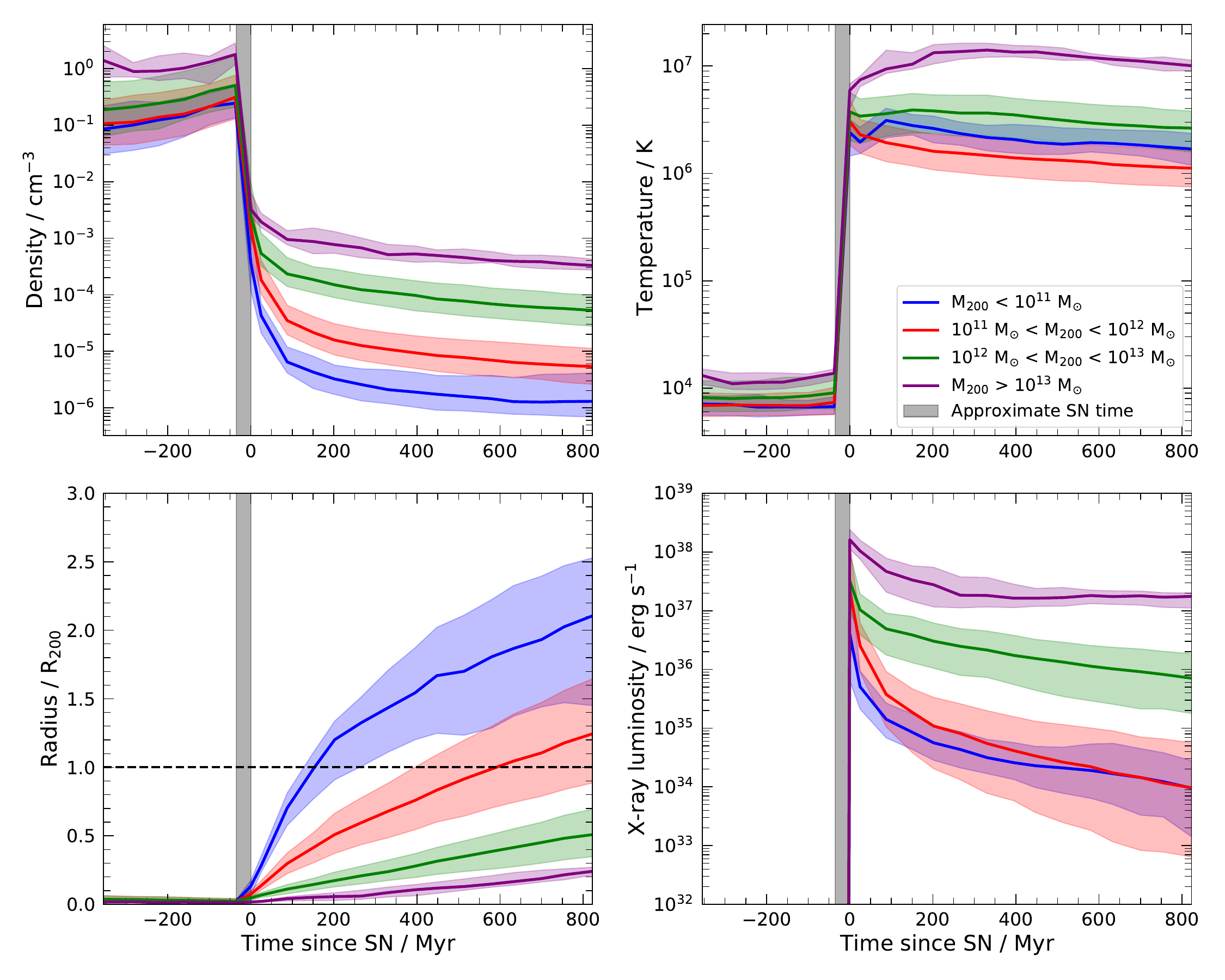}
    \caption{The properties of gas particles within the halo virial
      radius of disc galaxies in the EAGLE reference simulation, which
      have been subject to direct SNe energy feedback within the last
      $25 \units{Myr}$ of $z = 0.1$. The particles are separated
      into categories based on the $z = 0$ virial mass, $M_{200}$, of
      the host halo. We show the median property, at each time output,
      for all particles in halo mass ranges:
      $M_{200} < 10^{11} \units{\msun}$,
      $10^{11} < M_{200}/\units{\msun} < 10^{12}$,
      $10^{12} < M_{200}/\units{\msun} < 10^{13}$ and
      $M_{200} > 10^{13} \units{\msun}$, shown by the blue, red, green
      and purple lines respectively. The median atomic number density,
      temperature, radius and X-ray luminosity, of each particle are
      show in the upper-left, upper-right, bottom-left and
      bottom-right panels respectively. The shaded regions enclose the
      $20$th and $80$th percentiles.}
    \label{fig:missing_feedback}
\end{figure*}

The inner region of the hot X-ray-emitting coronae around disc
galaxies is the site of a complex interplay between the accreted, 
quasi-hydrostatic halo and the hot, metal-enriched winds driven by
SNe feedback within the ISM \citep{putman:2012}. However, X-ray
observations of the inner regions of star-forming disc galaxies
typically find relatively low X-ray luminosities \citep{li:2012}. The
observed X-ray luminosities can be compared with the rate of energy
input into galaxies from SNe feedback. The coupling efficiency,
$\eta_{X}$, is defined as the ratio of the observed X-ray luminosity
to the rate of energy input from SNe which can be calculated from the
inferred recent star formation rate,
$\eta_{X} = L_{\mathrm{X}} / \dot{E}_{\mathrm{SN}}$. In real observations, the
mean value of this coupling efficiency has been found to be very
small, approximately $\eta_{X} \approx 0.004$ \citep{li:2013}, and
thus the energy input by SNe is said to be ``missing''. The apparent
low X-ray luminosity is sometimes referred to as the ``missing
feedback'' problem \citep[e.g.][]{wang:2010}.

The fate of SNe-heated gas falls into one of three categories:
\begin{enumerate}
    \item \textbf{halo ejection:} hot gas is rapidly blown out of the galactic halo and joins the intergalactic medium;
    \item \textbf{galaxy ejection:} hot gas is ejected from the galaxy but remains within the virial radius of the halo; 
    \item \textbf{galactic fountain:} outflowing gas cools rapidly and infalls back into the ISM of the galaxy.
\end{enumerate}

The low X-ray luminosity observed within the central regions of
star-forming galaxies can be used to distinguish amongst these three
possibilities.  For example, in the case of an efficient galactic
fountain, it is expected that there will be significant X-ray emission
concentrated around the central regions of the halo. These X-rays will
be emitted by hot, dense gas as it cools and falls back into the
galaxy. The other two scenarios will lead to hot gas moving outwards through the galactic halo and thus reducing in density. These halos
will have much less X-ray emission, for the same SFR, as less energy
is radiated due to the lower density.

The value of the X-ray coupling efficiency in the EAGLE reference
simulation is shown in Fig.~\ref{fig:eta_m200} for a sample of
isolated disc galaxies. We calculate the X-ray luminosity for all gas
particles within a sphere of radius, $R_{200}$, centred on the centre
of mass of the halo. The rate of energy injection by SNe,
$\dot{E}_{\mathrm{SN}}$, is known in the EAGLE simulations: the
subgrid model assumes that, on average,
$8.73 \times 10^{15} \units{erg~g^{-1}}$ of energy is injected per
unit of initial stellar mass formed. This can be combined with the
mean star formation within the last $250 \units{Myr}$ to calculate the
mean energy injection rate.

Fig.~\ref{fig:eta_m200} shows that, typically, the X-ray coupling
efficiency increases with halo mass. This is consistent with the work
of \cite{li:2013} who suggest that halos with more gas in their
central regions are able to retain more of the gas heated by feedback
and thus be more X-ray luminous at a given SFR. The correlation
between halo mass and X-ray coupling efficiency is consistent with the
results of Fig.~\ref{fig:eagle_scaling} which shows that in very low
mass halos there is a very rapid increase in the X-ray luminosity with
halo mass. As we can see in Fig.~\ref{fig:eagle_scaling_prist_wind},
low mass halos of $\mvir \leq 10^{11.5} \units{\msun}$, are
\textit{wind}-dominated and therefore the X-ray emission from them is
powered by SNe. In these halos, there is a positive correlation
between SFR and halo mass. As the X-ray coupling efficiency is also
proportional to the halo mass, this results in a steep relationship
between X-ray luminosity and halo mass for low mass halos.

We now attempt to understand the origin of the varying X-ray coupling
efficiency in the EAGLE simulations. We start by selecting all gas
particles within the virial radius of disc galaxies in the EAGLE reference
simulation that have been subject to direct SNe heating within the
last $25 \units{Myr}$ at $z = 0.1$. We then use a series of
high-cadence outputs from the simulations to track several properties
of the selected particles as a function of time, starting
approximately $300 \units{Myr}$ before the feedback event, until
$\approx 800 \units{Myr}$ after. The particles are then separated into
four $z= 0$ host halo mass bins: $M_{200} < 10^{11} \units{\msun}$,
$10^{11} < M_{200}/\units{\msun} < 10^{12}$,
$10^{12} < M_{200}/\units{\msun} < 10^{13}$ and
$M_{200} > 10^{13} \units{\msun}$. We plot the median atomic number
density, temperature, radius and X-ray luminosity of each particle in
each halo mass bin at every output. These are shown in the upper-left,
upper-right, bottom-left and bottom-right panels of
Fig.~\ref{fig:missing_feedback} respectively. The shaded regions show
the bands that enclose the $20$th and $80$th percentiles.

In Fig.~\ref{fig:missing_feedback} we see that prior to the feedback
event, the majority of the gas is dense
($\approx 0.1 \units{cm^{-3}}$), cold ($\approx 10^{4} \units{K}$) and
near the centre of the halo. This is consistent with being the ISM of
the central galaxy. In the output immediately after the feedback
event, the gas undergoes an almost instantaneous temperature increase
to approximately $10^{7} \units{K}$, which is lower than the peak
temperature of $10^{7.5} \units{K}$ imposed by SNe feedback. This
suggests that the initial cooling rate must be very high and, as a
result, the maximum temperature is poorly sampled.  In the
$25 \units{Myr}$ after the feedback event, the median temperature of
the SNe-heated gas, in all but the most massive halos, declines to
$\approx (1 - 3) \times 10^{6} \units{K}$. As the gas evolves further,
the median temperature stabilises; this is likely because the density
has decreased and thus, the radiative cooling efficiency has also
decreased. As the density and temperature stabilise, the X-ray
luminosity reaches an approximately constant value that correlates
with halo mass. In the most massive halos,
$\geq 10^{13} \units{\msun}$, the post-SNe temperature of the gas is
approximately a constant value of $10^{7} \units{K}$. This high
temperature is likely the result of the feedback-heated gas
approaching thermal equilibrium as it mixes with the surrounding hot
halo gas.

The 3D physical distance from the centre of the galaxy, normalised by
the virial radius of the host halo, is plotted in the bottom-left
panel of Fig.~\ref{fig:missing_feedback} as a function of time since
the feedback event. In the lowest mass halos,
$M_{200} < 10^{11} \units{\msun}$, the median gas particle subject to
feedback leaves the virial radius of the halo in less than
$200 \units{Myr}$. On the other hand, in the more massive halos,
$10^{12} < M_{200}/\units{\msun} < 10^{13}$, the median particle is
still within the inner half of the virial radius after $500
\units{Myr}$. In lower mass halos, the density drops more
rapidly. These effects combine to compound the ``missing feedback''
problem in low mass halos, as seen in Fig.~\ref{fig:eta_m200}. It is
also likely that these effects are not independent, e.g. leaving the
halo more rapidly will result in faster decreasing densities.

Fundamentally, the EAGLE simulations predict that the so-called
``missing feedback'' is associated with outflowing material. Therefore
we predict that halos with galactic fountains should have much higher
X-ray coupling efficiencies. Future observations might be able to
distinguish between efficient galactic fountains, as in the Auriga
simulations \citep{grand:2019}, and the more outflowing,
baryon-deficient halos produced in the EAGLE simulations.

\section{Conclusions}\label{sec:conclusions}

The presence of hot accreted X-ray luminous gaseous atmospheres in
quasi-hydrostatic equilibrium around late-type galaxies in halos of
mass $\ge 10^{12} \units{\msun}$ is a fundamental prediction of galaxy
formation models within $\Lambda$CDM \citep{white:1991}. After
numerous failed searches for diffuse X-ray emission from these
galaxies over the years, detections were finally forthcoming in the
early 2000s, as discussed in the Introduction. However, these
detections have often been attributed to hot, outflowing winds driven
by feedback in the ISM of the galaxy rather than by accreted
coronae. We have used the large-volume, cosmological hydrodynamics
EAGLE simulations to investigate the origin and properties of the hot,
X-ray emitting gas around disc galaxies. Although our focus has been
on emission from individual $\sim L^\star$ galaxies, we have also
briefly considered more and less massive galaxies, particularly when
discussing scaling relations. Our main results are as follows:

\begin{itemize}
\item The EAGLE simulations predict that MW-mass halos are baryon
  deficient relative to the mean baryon faction (see
  Fig.~\ref{fig:eagle_fb}). Specifically, they contain only
  $\approx 40\%$ of the mean cosmic baryon fraction within their halo
  virial radius. About half of the baryons are present as a hot
  ($T > 10^{5.5} \units{K}$) gaseous halo. This baryon deficiency is
  attributed to feedback which can drive halo-wide winds. In more
  massive haloes, $M_{\rm vir} \geq 5 \times 10^{13} \units{\msun}$,
  the baryon fraction approaches the cosmic mean.
    
\item The central halo X-ray emission in the EAGLE simulations is
  dominated by winds triggered by SNe feedback. The X-ray emission in
  the outer regions is generally produced by an accreted,
  quasi-hydrostatic hot gaseous halo (see
  Fig.~\ref{fig:eagle_scaling_prist_wind}). Excising the inner
  $0.10 R_{\mathrm{vir}}$ of the halo is sufficient to remove the
  majority of the hot gas heated by SNe feedback in halos of mass
  $\geq 10^{12} \units{\msun}$, thus allowing us to probe the emission
  from hot, accreted atmospheres predicted in the analytic galaxy
  formation framework proposed by \cite{white:1978} and, for
  $\Lambda$CDM, by \cite{white:1991}.
    
\item The EAGLE simulations reproduce the observed general trend of
  X-ray luminosity with halo mass; however, the simulations typically
  overpredict the X-ray luminosity in the outer regions of haloes of
  mass $M_{\rm vir} \geq 2 \times 10^{12} \units{\msun}$ (see
  Fig.~\ref{fig:eagle_scaling}). The origin of the excess emission
  appears to be too high a gas fraction. Varying the parameters of AGN
  feedback to make it more expulsive, as in the AGNdT9-L050N0752
  simulation, can improve the agreement with the observed X-ray
  luminosities. In the inner regions, where winds dominate the X-ray
  luminosity, the EAGLE simulations reproduce both the trend and
  scatter of the observed \lxmvir\ relation.
    
\item We predict that any steepening of the logarithmic slope of the
  \lxmvir\ relation, above the self-similar value,
  $\alpha = 5/3$, is due to a varying hot gas fraction with halo mass. We show that the logarithmic slope of the power-law
  relationship between gas baryon fraction and halo mass can be inferred from the \lxmvir\ relation using an analytical relation we derived, Eqn. \ref{eq:lx_halo_mass_power}. This relation holds across several EAGLE simulations which employ different AGN feedback prescriptions. The same methodology can be applied to future X-ray
  observations to constrain the slope, normalisation and scatter of the halo gas fraction as a function of halo mass.
    
  \item We identify the physical origin of the so-called ``missing
    feedback'' around low-mass, star-forming galaxies. The EAGLE
    simulations suggest that much of the energy injected by SNe feedback is lost as hot gas is ejected from the halo into the low-density IGM (see Fig.~\ref{fig:missing_feedback}). Hot winds,
    driven by SNe feedback, can leave low mass halos of mass,
    $\approx 10^{11} \units{\msun}$, in a timescale of
    $\approx 100 \units{Myr}$. By contrast, in higher-mass halos,
    which have a more gas-rich central region, the outflowing gas is trapped at a higher density where it can radiate a larger fraction of the injected energy. This leads to a sharp increase in 
    X-ray luminosity as a function of halo mass within the central regions.
\end{itemize}

\section*{Acknowledgements}

AJK would like to thank Rob Crain and Jon Davies for a variety of engaging and informative discussions along with John Helly, Josh Borrow and Matthieu Schaller for their inexhaustible knowledge of all things EAGLE and hydrodynamics. We would also like to thank Joop Schaye and the anonymous referee for enlightening discussions and comments which significantly improved the manuscript. This work was supported by the Science and Technology Facilities Council (STFC) consolidated grant ST/P000541/1. AJK acknowledges an STFC studentship grant ST/S505365/1. CSF acknowledges support by the European Research Council (ERC) through Advanced Investigator grant DMIDAS (GA 786910). This work used the DiRAC@Durham facility managed by the Institute for Computational Cosmology on behalf of the STFC DiRAC HPC Facility (www.dirac.ac.uk). The equipment was funded by BEIS capital funding via STFC capital grants ST/K00042X/1, ST/P002293/1, ST/R002371/1 and ST/S002502/1, Durham University and STFC operations grant ST/R000832/1. DiRAC is part of the National e-Infrastructure.

\section*{Software Citations}

This paper made use of the following software packages:
\begin{itemize}
    \item Gadget \citep{springel:2005}

    \item {\tt python} \citep{python}, with the following libraries
    \begin{itemize}
        \item {\tt numpy} \citep{numpy}
        \item {\tt scipy} \citep{scipy}
        \item {\tt h5py} \citep{hdf5}
        \item {\tt matplotlib} \citep{mpl}
        \item {\tt numba} \citep{numba}
        \item {\tt mpi4py} \citep{mpi4py}
        \item {\tt unyt} \citep{unyt}
        \item {\tt pyatomdb} \citep{pyatomdb}
        \item {\tt read\_eagle} \citep{eagledb}
    \end{itemize}
\end{itemize}

\bibliographystyle{mnras}
\bibliography{ref}

\begin{thebibliography}{}
\makeatletter
\relax
\def\mn@urlcharsother{\let\do\@makeother \do\$\do\&\do\#\do\^\do\_\do\%\do\~}
\def\mn@doi{\begingroup\mn@urlcharsother \@ifnextchar [ {\mn@doi@}
  {\mn@doi@[]}}
\def\mn@doi@[#1]#2{\def\@tempa{#1}\ifx\@tempa\@empty \href
  {http://dx.doi.org/#2} {doi:#2}\else \href {http://dx.doi.org/#2} {#1}\fi
  \endgroup}
\def\mn@eprint#1#2{\mn@eprint@#1:#2::\@nil}
\def\mn@eprint@arXiv#1{\href {http://arxiv.org/abs/#1} {{\tt arXiv:#1}}}
\def\mn@eprint@dblp#1{\href {http://dblp.uni-trier.de/rec/bibtex/#1.xml}
  {dblp:#1}}
\def\mn@eprint@#1:#2:#3:#4\@nil{\def\@tempa {#1}\def\@tempb {#2}\def\@tempc
  {#3}\ifx \@tempc \@empty \let \@tempc \@tempb \let \@tempb \@tempa \fi \ifx
  \@tempb \@empty \def\@tempb {arXiv}\fi \@ifundefined
  {mn@eprint@\@tempb}{\@tempb:\@tempc}{\expandafter \expandafter \csname
  mn@eprint@\@tempb\endcsname \expandafter{\@tempc}}}

\bibitem[\protect\citeauthoryear{{Anders} \& {Grevesse}}{{Anders} \&
  {Grevesse}}{1989}]{anders:1989}
{Anders} E.,  {Grevesse} N.,  1989, \mn@doi [\gca]
  {10.1016/0016-7037(89)90286-X}, \href
  {http://adsabs.harvard.edu/abs/1989GeCoA..53..197A} {53, 197}

\bibitem[\protect\citeauthoryear{{Anderson}, {Gaspari}, {White}, {Wang}  \&
  {Dai}}{{Anderson} et~al.}{2015}]{anderson:2015}
{Anderson} M.~E.,  {Gaspari} M.,  {White} S. D.~M.,  {Wang} W.,   {Dai} X.,
  2015, \mn@doi [\mnras] {10.1093/mnras/stv437}, \href
  {https://ui.adsabs.harvard.edu/abs/2015MNRAS.449.3806A} {449, 3806}

\bibitem[\protect\citeauthoryear{{Anderson}, {Churazov}  \&
  {Bregman}}{{Anderson} et~al.}{2016}]{anderson:2016}
{Anderson} M.~E.,  {Churazov} E.,   {Bregman} J.~N.,  2016, \mn@doi [\mnras]
  {10.1093/mnras/stv2314}, \href
  {http://adsabs.harvard.edu/abs/2016MNRAS.455..227A} {455, 227}

\bibitem[\protect\citeauthoryear{{Bah{\'e}} et~al.,}{{Bah{\'e}}
  et~al.}{2016}]{bahe:2016}
{Bah{\'e}} Y.~M.,  et~al., 2016, \mn@doi [\mnras] {10.1093/mnras/stv2674},
  \href {https://ui.adsabs.harvard.edu/abs/2016MNRAS.456.1115B} {456, 1115}

\bibitem[\protect\citeauthoryear{{Behroozi}, {Wechsler}  \&
  {Conroy}}{{Behroozi} et~al.}{2013}]{behroozi:2013}
{Behroozi} P.~S.,  {Wechsler} R.~H.,   {Conroy} C.,  2013, \mn@doi [\apj]
  {10.1088/0004-637X/770/1/57}, \href
  {https://ui.adsabs.harvard.edu/abs/2013ApJ...770...57B} {770, 57}

\bibitem[\protect\citeauthoryear{{Benson}, {Bower}, {Frenk}  \&
  {White}}{{Benson} et~al.}{2000}]{benson:2000}
{Benson} A.~J.,  {Bower} R.~G.,  {Frenk} C.~S.,   {White} S.~D.~M.,  2000,
  \mn@doi [\mnras] {10.1046/j.1365-8711.2000.03362.x}, \href
  {http://adsabs.harvard.edu/abs/2000MNRAS.314..557B} {314, 557}

\bibitem[\protect\citeauthoryear{{Bogd{\'a}n}, {Forman}, {Kraft}  \&
  {Jones}}{{Bogd{\'a}n} et~al.}{2013}]{bogdan:2013}
{Bogd{\'a}n} {\'A}.,  {Forman} W.~R.,  {Kraft} R.~P.,   {Jones} C.,  2013,
  \mn@doi [\apj] {10.1088/0004-637X/772/2/98}, \href
  {http://adsabs.harvard.edu/abs/2013ApJ...772...98B} {772, 98}

\bibitem[\protect\citeauthoryear{{Bogd{\'a}n} et~al.,}{{Bogd{\'a}n}
  et~al.}{2015}]{bogdan:2015}
{Bogd{\'a}n} {\'A}.,  et~al., 2015, \mn@doi [\apj]
  {10.1088/0004-637X/804/1/72}, \href
  {https://ui.adsabs.harvard.edu/abs/2015ApJ...804...72B} {804, 72}

\bibitem[\protect\citeauthoryear{{Bogd{\'a}n}, {Bourdin}, {Forman}, {Kraft},
  {Vogelsberger}, {Hernquist}  \& {Springel}}{{Bogd{\'a}n}
  et~al.}{2017}]{bogdan:2017}
{Bogd{\'a}n} {\'A}.,  {Bourdin} H.,  {Forman} W.~R.,  {Kraft} R.~P.,
  {Vogelsberger} M.,  {Hernquist} L.,   {Springel} V.,  2017, \mn@doi [\apj]
  {10.3847/1538-4357/aa9523}, \href
  {http://adsabs.harvard.edu/abs/2017ApJ...850...98B} {850, 98}

\bibitem[\protect\citeauthoryear{{B{\"o}hringer}, {Dolag}  \&
  {Chon}}{{B{\"o}hringer} et~al.}{2012}]{bohringer:2012}
{B{\"o}hringer} H.,  {Dolag} K.,   {Chon} G.,  2012, \mn@doi [\aap]
  {10.1051/0004-6361/201118000}, \href
  {https://ui.adsabs.harvard.edu/abs/2012A&A...539A.120B} {539, A120}

\bibitem[\protect\citeauthoryear{{Bower}, {Schaye}, {Frenk}, {Theuns},
  {Schaller}, {Crain}  \& {McAlpine}}{{Bower} et~al.}{2017}]{bower:2017}
{Bower} R.~G.,  {Schaye} J.,  {Frenk} C.~S.,  {Theuns} T.,  {Schaller} M.,
  {Crain} R.~A.,   {McAlpine} S.,  2017, \mn@doi [\mnras]
  {10.1093/mnras/stw2735}, \href
  {https://ui.adsabs.harvard.edu/abs/2017MNRAS.465...32B} {465, 32}

\bibitem[\protect\citeauthoryear{{Bregman}, {Anderson}, {Miller},
  {Hodges-Kluck}, {Dai}, {Li}, {Li}  \& {Qu}}{{Bregman}
  et~al.}{2018}]{bregman:2018}
{Bregman} J.~N.,  {Anderson} M.~E.,  {Miller} M.~J.,  {Hodges-Kluck} E.,  {Dai}
  X.,  {Li} J.-T.,  {Li} Y.,   {Qu} Z.,  2018, \mn@doi [\apj]
  {10.3847/1538-4357/aacafe}, \href
  {https://ui.adsabs.harvard.edu/abs/2018ApJ...862....3B} {862, 3}

\bibitem[\protect\citeauthoryear{Collette}{Collette}{2013}]{hdf5}
Collette A.,  2013, Python and HDF5.
O'Reilly

\bibitem[\protect\citeauthoryear{{Correa}, {Schaye}, {van de Voort}, {Duffy}
  \& {Wyithe}}{{Correa} et~al.}{2018}]{correa:2018}
{Correa} C.~A.,  {Schaye} J.,  {van de Voort} F.,  {Duffy} A.~R.,   {Wyithe} J.
  S.~B.,  2018, \mn@doi [\mnras] {10.1093/mnras/sty871}, \href
  {https://ui.adsabs.harvard.edu/abs/2018MNRAS.478..255C} {478, 255}

\bibitem[\protect\citeauthoryear{{Crain} et~al.,}{{Crain}
  et~al.}{2009}]{crain:2009}
{Crain} R.~A.,  et~al., 2009, \mn@doi [\mnras]
  {10.1111/j.1365-2966.2009.15402.x}, \href
  {http://adsabs.harvard.edu/abs/2009MNRAS.399.1773C} {399, 1773}

\bibitem[\protect\citeauthoryear{{Crain}, {McCarthy}, {Frenk}, {Theuns}  \&
  {Schaye}}{{Crain} et~al.}{2010}]{crain:2010}
{Crain} R.~A.,  {McCarthy} I.~G.,  {Frenk} C.~S.,  {Theuns} T.,   {Schaye} J.,
  2010, \mn@doi [\mnras] {10.1111/j.1365-2966.2010.16985.x}, \href
  {https://ui.adsabs.harvard.edu/abs/2010MNRAS.407.1403C} {407, 1403}

\bibitem[\protect\citeauthoryear{{Crain} et~al.,}{{Crain}
  et~al.}{2015}]{crain:2015}
{Crain} R.~A.,  et~al., 2015, \mn@doi [\mnras] {10.1093/mnras/stv725}, \href
  {https://ui.adsabs.harvard.edu/abs/2015MNRAS.450.1937C} {450, 1937}

\bibitem[\protect\citeauthoryear{{Creasey}, {Theuns}, {Bower}  \&
  {Lacey}}{{Creasey} et~al.}{2011}]{creasey:2011}
{Creasey} P.,  {Theuns} T.,  {Bower} R.~G.,   {Lacey} C.~G.,  2011, \mn@doi
  [\mnras] {10.1111/j.1365-2966.2011.19001.x}, \href
  {https://ui.adsabs.harvard.edu/abs/2011MNRAS.415.3706C} {415, 3706}

\bibitem[\protect\citeauthoryear{{Cullen} \& {Dehnen}}{{Cullen} \&
  {Dehnen}}{2010}]{cullen:2010}
{Cullen} L.,  {Dehnen} W.,  2010, \mn@doi [\mnras]
  {10.1111/j.1365-2966.2010.17158.x}, \href
  {https://ui.adsabs.harvard.edu/abs/2010MNRAS.408..669C} {408, 669}

\bibitem[\protect\citeauthoryear{Dalcin, Paz, Kler  \& Cosimo}{Dalcin
  et~al.}{2011}]{mpi4py}
Dalcin L.~D.,  Paz R.~R.,  Kler P.~A.,   Cosimo A.,  2011, \mn@doi [Advances in
  Water Resources] {https://doi.org/10.1016/j.advwatres.2011.04.013}, 34, 1124

\bibitem[\protect\citeauthoryear{{Dalla Vecchia} \& {Schaye}}{{Dalla Vecchia}
  \& {Schaye}}{2012}]{vecchia:2012}
{Dalla Vecchia} C.,  {Schaye} J.,  2012, \mn@doi [\mnras]
  {10.1111/j.1365-2966.2012.21704.x}, \href
  {https://ui.adsabs.harvard.edu/abs/2012MNRAS.426..140D} {426, 140}

\bibitem[\protect\citeauthoryear{{Davies}, {Crain}, {McCarthy}, {Oppenheimer},
  {Schaye}, {Schaller}  \& {McAlpine}}{{Davies} et~al.}{2019}]{davies:2019}
{Davies} J.~J.,  {Crain} R.~A.,  {McCarthy} I.~G.,  {Oppenheimer} B.~D.,
  {Schaye} J.,  {Schaller} M.,   {McAlpine} S.,  2019, \mn@doi [\mnras]
  {10.1093/mnras/stz635}, \href
  {https://ui.adsabs.harvard.edu/abs/2019MNRAS.485.3783D} {485, 3783}

\bibitem[\protect\citeauthoryear{{Davis}, {Efstathiou}, {Frenk}  \&
  {White}}{{Davis} et~al.}{1985}]{davis:1985}
{Davis} M.,  {Efstathiou} G.,  {Frenk} C.~S.,   {White} S.~D.~M.,  1985,
  \mn@doi [\apj] {10.1086/163168}, \href
  {https://ui.adsabs.harvard.edu/abs/1985ApJ...292..371D} {292, 371}

\bibitem[\protect\citeauthoryear{{Dolag}, {Borgani}, {Murante}  \&
  {Springel}}{{Dolag} et~al.}{2009}]{dolag:2009}
{Dolag} K.,  {Borgani} S.,  {Murante} G.,   {Springel} V.,  2009, \mn@doi
  [\mnras] {10.1111/j.1365-2966.2009.15034.x}, \href
  {https://ui.adsabs.harvard.edu/abs/2009MNRAS.399..497D} {399, 497}

\bibitem[\protect\citeauthoryear{{Faerman}, {Sternberg}  \& {McKee}}{{Faerman}
  et~al.}{2017}]{faerman:2017}
{Faerman} Y.,  {Sternberg} A.,   {McKee} C.~F.,  2017, \mn@doi [\apj]
  {10.3847/1538-4357/835/1/52}, \href
  {https://ui.adsabs.harvard.edu/abs/2017ApJ...835...52F} {835, 52}

\bibitem[\protect\citeauthoryear{{Foster}, {Ji}, {Smith}  \&
  {Brickhouse}}{{Foster} et~al.}{2012}]{foster:2012}
{Foster} A.~R.,  {Ji} L.,  {Smith} R.~K.,   {Brickhouse} N.~S.,  2012, \mn@doi
  [\apj] {10.1088/0004-637X/756/2/128}, \href
  {https://ui.adsabs.harvard.edu/#abs/2012ApJ...756..128F} {756, 128}

\bibitem[\protect\citeauthoryear{{Foster}, {Smith}, {Brickhouse}  \&
  {Cui}}{{Foster} et~al.}{2016}]{pyatomdb}
{Foster} A.,  {Smith} R.~K.,  {Brickhouse} N.~S.,   {Cui} X.,  2016, in
  American Astronomical Society Meeting Abstracts \#227. p. 211.08

\bibitem[\protect\citeauthoryear{{Furlong} et~al.,}{{Furlong}
  et~al.}{2015}]{furlong:2015}
{Furlong} M.,  et~al., 2015, \mn@doi [\mnras] {10.1093/mnras/stv852}, \href
  {https://ui.adsabs.harvard.edu/abs/2015MNRAS.450.4486F} {450, 4486}

\bibitem[\protect\citeauthoryear{{Furlong} et~al.,}{{Furlong}
  et~al.}{2017}]{furlong:2017}
{Furlong} M.,  et~al., 2017, \mn@doi [\mnras] {10.1093/mnras/stw2740}, \href
  {https://ui.adsabs.harvard.edu/abs/2017MNRAS.465..722F} {465, 722}

\bibitem[\protect\citeauthoryear{{Gingold} \& {Monaghan}}{{Gingold} \&
  {Monaghan}}{1977}]{gingold:1977}
{Gingold} R.~A.,  {Monaghan} J.~J.,  1977, \mn@doi [\mnras]
  {10.1093/mnras/181.3.375}, \href
  {https://ui.adsabs.harvard.edu/abs/1977MNRAS.181..375G} {181, 375}

\bibitem[\protect\citeauthoryear{Goldbaum, ZuHone, Turk, Kowalik  \&
  Rosen}{Goldbaum et~al.}{2018}]{unyt}
Goldbaum N.~J.,  ZuHone J.~A.,  Turk M.~J.,  Kowalik K.,   Rosen A.~L.,  2018,
  \mn@doi [Journal of Open Source Software] {10.21105/joss.00809}, 3, 809

\bibitem[\protect\citeauthoryear{{Grand} et~al.,}{{Grand}
  et~al.}{2019}]{grand:2019}
{Grand} R. J.~J.,  et~al., 2019, \mn@doi [\mnras] {10.1093/mnras/stz2928},
  \href {https://ui.adsabs.harvard.edu/abs/2019MNRAS.490.4786G} {490, 4786}

\bibitem[\protect\citeauthoryear{{Gupta}, {Mathur}, {Krongold}, {Nicastro}  \&
  {Galeazzi}}{{Gupta} et~al.}{2012}]{gupta:2012}
{Gupta} A.,  {Mathur} S.,  {Krongold} Y.,  {Nicastro} F.,   {Galeazzi} M.,
  2012, \mn@doi [\apjl] {10.1088/2041-8205/756/1/L8}, \href
  {https://ui.adsabs.harvard.edu/abs/2012ApJ...756L...8G} {756, L8}

\bibitem[\protect\citeauthoryear{{Hernquist} \& {Springel}}{{Hernquist} \&
  {Springel}}{2003}]{hernquist:2003}
{Hernquist} L.,  {Springel} V.,  2003, \mn@doi [\mnras]
  {10.1046/j.1365-8711.2003.06499.x}, \href
  {https://ui.adsabs.harvard.edu/abs/2003MNRAS.341.1253H} {341, 1253}

\bibitem[\protect\citeauthoryear{{Hodges-Kluck}, {Bregman}  \&
  {Li}}{{Hodges-Kluck} et~al.}{2018}]{hodges-kluck:2018}
{Hodges-Kluck} E.~J.,  {Bregman} J.~N.,   {Li} J.-t.,  2018, \mn@doi [\apj]
  {10.3847/1538-4357/aae38a}, \href
  {http://adsabs.harvard.edu/abs/2018ApJ...866..126H} {866, 126}

\bibitem[\protect\citeauthoryear{{Hopkins}}{{Hopkins}}{2013}]{hopkins:2013}
{Hopkins} P.~F.,  2013, \mn@doi [\mnras] {10.1093/mnras/sts210}, \href
  {https://ui.adsabs.harvard.edu/abs/2013MNRAS.428.2840H} {428, 2840}

\bibitem[\protect\citeauthoryear{Hunter}{Hunter}{2007}]{mpl}
Hunter J.~D.,  2007, \mn@doi [Computing in Science \& Engineering]
  {10.1109/MCSE.2007.55}, 9, 90

\bibitem[\protect\citeauthoryear{Jones, Oliphant, Peterson  et~al.}{Jones
  et~al.}{2001}]{scipy}
Jones E.,  Oliphant T.,  Peterson P.,   et~al., 2001, {SciPy}: Open source
  scientific tools for {Python}, \url {http://www.scipy.org/}

\bibitem[\protect\citeauthoryear{{Kaiser}}{{Kaiser}}{1986}]{kaiser:1986}
{Kaiser} N.,  1986, \mn@doi [\mnras] {10.1093/mnras/222.2.323}, \href
  {https://ui.adsabs.harvard.edu/abs/1986MNRAS.222..323K} {222, 323}

\bibitem[\protect\citeauthoryear{{Kauffmann} et~al.,}{{Kauffmann}
  et~al.}{2003}]{kauffmann:2003}
{Kauffmann} G.,  et~al., 2003, \mn@doi [\mnras]
  {10.1111/j.1365-2966.2003.07154.x}, \href
  {https://ui.adsabs.harvard.edu/abs/2003MNRAS.346.1055K} {346, 1055}

\bibitem[\protect\citeauthoryear{{Keller}, {Wadsley}, {Benincasa}  \&
  {Couchman}}{{Keller} et~al.}{2014}]{keller:2014}
{Keller} B.~W.,  {Wadsley} J.,  {Benincasa} S.~M.,   {Couchman} H.~M.~P.,
  2014, \mn@doi [\mnras] {10.1093/mnras/stu1058}, \href
  {https://ui.adsabs.harvard.edu/abs/2014MNRAS.442.3013K} {442, 3013}

\bibitem[\protect\citeauthoryear{{Lacey} \& {Cole}}{{Lacey} \&
  {Cole}}{1994}]{lacey:1994}
{Lacey} C.,  {Cole} S.,  1994, \mn@doi [\mnras] {10.1093/mnras/271.3.676},
  \href {https://ui.adsabs.harvard.edu/abs/1994MNRAS.271..676L} {271, 676}

\bibitem[\protect\citeauthoryear{{Lagos} et~al.,}{{Lagos}
  et~al.}{2015}]{lagos:2015}
{Lagos} C. d.~P.,  et~al., 2015, \mn@doi [\mnras] {10.1093/mnras/stv1488},
  \href {https://ui.adsabs.harvard.edu/abs/2015MNRAS.452.3815L} {452, 3815}

\bibitem[\protect\citeauthoryear{Lam, Pitrou  \& Seibert}{Lam
  et~al.}{2015}]{numba}
Lam S.~K.,  Pitrou A.,   Seibert S.,  2015, in Proceedings of the Second
  Workshop on the LLVM Compiler Infrastructure in HPC. LLVM '15.
ACM, New York, NY, USA, pp 7:1--7:6, \mn@doi{10.1145/2833157.2833162}, \url
  {http://doi.acm.org/10.1145/2833157.2833162}

\bibitem[\protect\citeauthoryear{{Larson}}{{Larson}}{1974}]{larson:1974}
{Larson} R.~B.,  1974, \mn@doi [\mnras] {10.1093/mnras/166.3.585}, \href
  {https://ui.adsabs.harvard.edu/abs/1974MNRAS.166..585L} {166, 585}

\bibitem[\protect\citeauthoryear{{Li} \& {Wang}}{{Li} \&
  {Wang}}{2013a}]{li:2012}
{Li} J.-T.,  {Wang} Q.~D.,  2013a, \mn@doi [\mnras] {10.1093/mnras/sts183},
  \href {https://ui.adsabs.harvard.edu/\#abs/2013MNRAS.428.2085L} {428, 2085}

\bibitem[\protect\citeauthoryear{{Li} \& {Wang}}{{Li} \&
  {Wang}}{2013b}]{li:2013}
{Li} J.-T.,  {Wang} Q.~D.,  2013b, \mn@doi [\mnras] {10.1093/mnras/stt1501},
  \href {https://ui.adsabs.harvard.edu/\#abs/2013MNRAS.435.3071L} {435, 3071}

\bibitem[\protect\citeauthoryear{{Li}, {Wang}  \& {Hameed}}{{Li}
  et~al.}{2007}]{li:2007}
{Li} Z.,  {Wang} Q.~D.,   {Hameed} S.,  2007, \mn@doi [\mnras]
  {10.1111/j.1365-2966.2007.11513.x}, \href
  {https://ui.adsabs.harvard.edu/abs/2007MNRAS.376..960L} {376, 960}

\bibitem[\protect\citeauthoryear{{Li}, {Crain}  \& {Wang}}{{Li}
  et~al.}{2014}]{li:2014}
{Li} J.-T.,  {Crain} R.~A.,   {Wang} Q.~D.,  2014, \mn@doi [\mnras]
  {10.1093/mnras/stu329}, \href
  {http://adsabs.harvard.edu/abs/2014MNRAS.440..859L} {440, 859}

\bibitem[\protect\citeauthoryear{{Li}, {Bregman}, {Wang}, {Crain}, {Anderson}
  \& {Zhang}}{{Li} et~al.}{2017}]{li:2017}
{Li} J.-T.,  {Bregman} J.~N.,  {Wang} Q.~D.,  {Crain} R.~A.,  {Anderson} M.~E.,
    {Zhang} S.,  2017, \mn@doi [\apjs] {10.3847/1538-4365/aa96fc}, \href
  {http://adsabs.harvard.edu/abs/2017ApJS..233...20L} {233, 20}

\bibitem[\protect\citeauthoryear{{Lin}, {Stanford}, {Eisenhardt}, {Vikhlinin},
  {Maughan}  \& {Kravtsov}}{{Lin} et~al.}{2012}]{lin:2012}
{Lin} Y.-T.,  {Stanford} S.~A.,  {Eisenhardt} P. R.~M.,  {Vikhlinin} A.,
  {Maughan} B.~J.,   {Kravtsov} A.,  2012, \mn@doi [\apjl]
  {10.1088/2041-8205/745/1/L3}, \href
  {https://ui.adsabs.harvard.edu/abs/2012ApJ...745L...3L} {745, L3}

\bibitem[\protect\citeauthoryear{{Lucy}}{{Lucy}}{1977}]{lucy:1977}
{Lucy} L.~B.,  1977, \mn@doi [\aj] {10.1086/112164}, \href
  {https://ui.adsabs.harvard.edu/abs/1977AJ.....82.1013L} {82, 1013}

\bibitem[\protect\citeauthoryear{{McAlpine} et~al.,}{{McAlpine}
  et~al.}{2016}]{mcalpine:2016}
{McAlpine} S.,  et~al., 2016, \mn@doi [Astronomy and Computing]
  {10.1016/j.ascom.2016.02.004}, \href
  {https://ui.adsabs.harvard.edu/abs/2016A&C....15...72M} {15, 72}

\bibitem[\protect\citeauthoryear{{Mitchell}, {Schaye}, {Bower}  \&
  {Crain}}{{Mitchell} et~al.}{2019}]{mitchell:2019}
{Mitchell} P.~D.,  {Schaye} J.,  {Bower} R.~G.,   {Crain} R.~A.,  2019, arXiv
  e-prints, \href {https://ui.adsabs.harvard.edu/abs/2019arXiv191009566M} {p.
  arXiv:1910.09566}

\bibitem[\protect\citeauthoryear{{Moster}, {Naab}  \& {White}}{{Moster}
  et~al.}{2013}]{moster:2013}
{Moster} B.~P.,  {Naab} T.,   {White} S. D.~M.,  2013, \mn@doi [\mnras]
  {10.1093/mnras/sts261}, \href
  {https://ui.adsabs.harvard.edu/abs/2013MNRAS.428.3121M} {428, 3121}

\bibitem[\protect\citeauthoryear{{Navarro} \& {White}}{{Navarro} \&
  {White}}{1993}]{navarro:1993}
{Navarro} J.~F.,  {White} S.~D.~M.,  1993, \mn@doi [\mnras]
  {10.1093/mnras/265.2.271}, \href
  {https://ui.adsabs.harvard.edu/abs/1993MNRAS.265..271N} {265, 271}

\bibitem[\protect\citeauthoryear{{Navarro}, {Frenk}  \& {White}}{{Navarro}
  et~al.}{1997}]{navarro:1997}
{Navarro} J.~F.,  {Frenk} C.~S.,   {White} S.~D.~M.,  1997, \mn@doi [\apj]
  {10.1086/304888}, \href
  {https://ui.adsabs.harvard.edu/abs/1997ApJ...490..493N} {490, 493}

\bibitem[\protect\citeauthoryear{{Nicastro}, {Senatore}, {Gupta}, {Guainazzi},
  {Mathur}, {Krongold}, {Elvis}  \& {Piro}}{{Nicastro}
  et~al.}{2016}]{nicastro:2016}
{Nicastro} F.,  {Senatore} F.,  {Gupta} A.,  {Guainazzi} M.,  {Mathur} S.,
  {Krongold} Y.,  {Elvis} M.,   {Piro} L.,  2016, \mn@doi [\mnras]
  {10.1093/mnras/stv2923}, \href
  {https://ui.adsabs.harvard.edu/abs/2016MNRAS.457..676N} {457, 676}

\bibitem[\protect\citeauthoryear{{Oppenheimer} et~al.,}{{Oppenheimer}
  et~al.}{2020}]{oppenheimer:2020}
{Oppenheimer} B.~D.,  et~al., 2020, \mn@doi [\apjl] {10.3847/2041-8213/ab846f},
  \href {https://ui.adsabs.harvard.edu/abs/2020ApJ...893L..24O} {893, L24}

\bibitem[\protect\citeauthoryear{{Owen} \& {Warwick}}{{Owen} \&
  {Warwick}}{2009}]{owen:2009}
{Owen} R.~A.,  {Warwick} R.~S.,  2009, \mn@doi [\mnras]
  {10.1111/j.1365-2966.2009.14464.x}, \href
  {https://ui.adsabs.harvard.edu/abs/2009MNRAS.394.1741O} {394, 1741}

\bibitem[\protect\citeauthoryear{{Planck Collaboration} et~al.,}{{Planck
  Collaboration} et~al.}{2013}]{planck:2013}
{Planck Collaboration} et~al., 2013, \mn@doi [\aap]
  {10.1051/0004-6361/201220941}, \href
  {https://ui.adsabs.harvard.edu/abs/2013A&A...557A..52P} {557, A52}

\bibitem[\protect\citeauthoryear{{Pratt}, {Croston}, {Arnaud}  \&
  {B{\"o}hringer}}{{Pratt} et~al.}{2009}]{pratt:2009}
{Pratt} G.~W.,  {Croston} J.~H.,  {Arnaud} M.,   {B{\"o}hringer} H.,  2009,
  \mn@doi [\aap] {10.1051/0004-6361/200810994}, \href
  {https://ui.adsabs.harvard.edu/abs/2009A&A...498..361P} {498, 361}

\bibitem[\protect\citeauthoryear{{Price}}{{Price}}{2008}]{price:2008}
{Price} D.~J.,  2008, \mn@doi [Journal of Computational Physics]
  {10.1016/j.jcp.2008.08.011}, \href
  {https://ui.adsabs.harvard.edu/abs/2008JCoPh.22710040P} {227, 10040}

\bibitem[\protect\citeauthoryear{{Putman}, {Peek}  \& {Joung}}{{Putman}
  et~al.}{2012}]{putman:2012}
{Putman} M.~E.,  {Peek} J.~E.~G.,   {Joung} M.~R.,  2012, \mn@doi [\araa]
  {10.1146/annurev-astro-081811-125612}, \href
  {http://adsabs.harvard.edu/abs/2012ARA%26A..50..491P} {50, 491}

\bibitem[\protect\citeauthoryear{{Rosas-Guevara} et~al.,}{{Rosas-Guevara}
  et~al.}{2015}]{guevara:2015}
{Rosas-Guevara} Y.~M.,  et~al., 2015, \mn@doi [\mnras] {10.1093/mnras/stv2056},
  \href {https://ui.adsabs.harvard.edu/abs/2015MNRAS.454.1038R} {454, 1038}

\bibitem[\protect\citeauthoryear{{Rosas-Guevara}, {Bower}, {Schaye},
  {McAlpine}, {Dalla Vecchia}, {Frenk}, {Schaller}  \&
  {Theuns}}{{Rosas-Guevara} et~al.}{2016}]{guevara:2016}
{Rosas-Guevara} Y.,  {Bower} R.~G.,  {Schaye} J.,  {McAlpine} S.,  {Dalla
  Vecchia} C.,  {Frenk} C.~S.,  {Schaller} M.,   {Theuns} T.,  2016, \mn@doi
  [\mnras] {10.1093/mnras/stw1679}, \href
  {https://ui.adsabs.harvard.edu/abs/2016MNRAS.462..190R} {462, 190}

\bibitem[\protect\citeauthoryear{Rupke}{Rupke}{2018}]{rupke:2018}
Rupke D.,  2018, \mn@doi [Galaxies] {10.3390/galaxies6040138}, 6, 138

\bibitem[\protect\citeauthoryear{{Sales}, {Navarro}, {Theuns}, {Schaye},
  {White}, {Frenk}, {Crain}  \& {Dalla Vecchia}}{{Sales}
  et~al.}{2012}]{sales:2012}
{Sales} L.~V.,  {Navarro} J.~F.,  {Theuns} T.,  {Schaye} J.,  {White} S. D.~M.,
   {Frenk} C.~S.,  {Crain} R.~A.,   {Dalla Vecchia} C.,  2012, \mn@doi [\mnras]
  {10.1111/j.1365-2966.2012.20975.x}, \href
  {https://ui.adsabs.harvard.edu/abs/2012MNRAS.423.1544S} {423, 1544}

\bibitem[\protect\citeauthoryear{{Sarazin}}{{Sarazin}}{1986}]{sarazin:1986}
{Sarazin} C.~L.,  1986, \mn@doi [Reviews of Modern Physics]
  {10.1103/RevModPhys.58.1}, \href
  {https://ui.adsabs.harvard.edu/abs/1986RvMP...58....1S} {58, 1}

\bibitem[\protect\citeauthoryear{{Schaller} et~al.,}{{Schaller}
  et~al.}{2015}]{schaller:2015}
{Schaller} M.,  et~al., 2015, \mn@doi [\mnras] {10.1093/mnras/stv1067}, \href
  {https://ui.adsabs.harvard.edu/abs/2015MNRAS.451.1247S} {451, 1247}

\bibitem[\protect\citeauthoryear{{Schaye} \& {Dalla Vecchia}}{{Schaye} \&
  {Dalla Vecchia}}{2008}]{schaye:2008}
{Schaye} J.,  {Dalla Vecchia} C.,  2008, \mn@doi [\mnras]
  {10.1111/j.1365-2966.2007.12639.x}, \href
  {https://ui.adsabs.harvard.edu/abs/2008MNRAS.383.1210S} {383, 1210}

\bibitem[\protect\citeauthoryear{{Schaye} et~al.,}{{Schaye}
  et~al.}{2015}]{schaye:2015}
{Schaye} J.,  et~al., 2015, \mn@doi [\mnras] {10.1093/mnras/stu2058}, \href
  {http://adsabs.harvard.edu/abs/2015MNRAS.446..521S} {446, 521}

\bibitem[\protect\citeauthoryear{{Smith}, {Brickhouse}, {Liedahl}  \&
  {Raymond}}{{Smith} et~al.}{2001}]{smith:2001}
{Smith} R.~K.,  {Brickhouse} N.~S.,  {Liedahl} D.~A.,   {Raymond} J.~C.,  2001,
  \mn@doi [\apjl] {10.1086/322992}, \href
  {https://ui.adsabs.harvard.edu/abs/2001ApJ...556L..91S} {556, L91}

\bibitem[\protect\citeauthoryear{{Snowden} et~al.,}{{Snowden}
  et~al.}{1997}]{snowden:1997}
{Snowden} S.~L.,  et~al., 1997, \mn@doi [\apj] {10.1086/304399}, \href
  {https://ui.adsabs.harvard.edu/abs/1997ApJ...485..125S} {485, 125}

\bibitem[\protect\citeauthoryear{{Spitzer}}{{Spitzer}}{1956}]{spitzer:1956}
{Spitzer} Jr. L.,  1956, \mn@doi [\apj] {10.1086/146200}, \href
  {https://ui.adsabs.harvard.edu/abs/1956ApJ...124...20S} {124, 20}

\bibitem[\protect\citeauthoryear{{Springel}}{{Springel}}{2005}]{springel:2005}
{Springel} V.,  2005, \mn@doi [\mnras] {10.1111/j.1365-2966.2005.09655.x},
  \href {https://ui.adsabs.harvard.edu/abs/2005MNRAS.364.1105S} {364, 1105}

\bibitem[\protect\citeauthoryear{{Springel}, {White}, {Tormen}  \&
  {Kauffmann}}{{Springel} et~al.}{2001}]{springel:2001}
{Springel} V.,  {White} S.~D.~M.,  {Tormen} G.,   {Kauffmann} G.,  2001,
  \mn@doi [\mnras] {10.1046/j.1365-8711.2001.04912.x}, \href
  {http://adsabs.harvard.edu/abs/2001MNRAS.328..726S} {328, 726}

\bibitem[\protect\citeauthoryear{Strickland \& Heckman}{Strickland \&
  Heckman}{2007}]{strickland:2007}
Strickland D.~K.,  Heckman T.~M.,  2007, \mn@doi [The Astrophysical Journal]
  {10.1086/511174}, 658, 258–281

\bibitem[\protect\citeauthoryear{{Strickland}, {Heckman}, {Colbert}, {Hoopes}
  \& {Weaver}}{{Strickland} et~al.}{2004}]{strickland:2004}
{Strickland} D.~K.,  {Heckman} T.~M.,  {Colbert} E.~J.~M.,  {Hoopes} C.~G.,
  {Weaver} K.~A.,  2004, \mn@doi [\apjs] {10.1086/382214}, \href
  {http://adsabs.harvard.edu/abs/2004ApJS..151..193S} {151, 193}

\bibitem[\protect\citeauthoryear{{Sun}, {Voit}, {Donahue}, {Jones}, {Forman}
  \& {Vikhlinin}}{{Sun} et~al.}{2009}]{sun:2009}
{Sun} M.,  {Voit} G.~M.,  {Donahue} M.,  {Jones} C.,  {Forman} W.,
  {Vikhlinin} A.,  2009, \mn@doi [\apj] {10.1088/0004-637X/693/2/1142}, \href
  {https://ui.adsabs.harvard.edu/abs/2009ApJ...693.1142S} {693, 1142}

\bibitem[\protect\citeauthoryear{{Sunyaev} \& {Zeldovich}}{{Sunyaev} \&
  {Zeldovich}}{1970}]{sunyaev:1970}
{Sunyaev} R.~A.,  {Zeldovich} Y.~B.,  1970, \mn@doi [\apss]
  {10.1007/BF00653471}, \href
  {https://ui.adsabs.harvard.edu/abs/1970Ap&SS...7....3S} {7, 3}

\bibitem[\protect\citeauthoryear{{The EAGLE team}}{{The EAGLE
  team}}{2017}]{eagledb}
{The EAGLE team} 2017, arXiv e-prints, \href
  {https://ui.adsabs.harvard.edu/abs/2017arXiv170609899T} {p. arXiv:1706.09899}

\bibitem[\protect\citeauthoryear{{Trayford} et~al.,}{{Trayford}
  et~al.}{2015}]{trayford:2015}
{Trayford} J.~W.,  et~al., 2015, \mn@doi [\mnras] {10.1093/mnras/stv1461},
  \href {https://ui.adsabs.harvard.edu/abs/2015MNRAS.452.2879T} {452, 2879}

\bibitem[\protect\citeauthoryear{{Trayford} et~al.,}{{Trayford}
  et~al.}{2017}]{trayford:2017}
{Trayford} J.~W.,  et~al., 2017, \mn@doi [\mnras] {10.1093/mnras/stx1051},
  \href {https://ui.adsabs.harvard.edu/abs/2017MNRAS.470..771T} {470, 771}

\bibitem[\protect\citeauthoryear{{T{\"u}llmann}, {Pietsch}, {Rossa},
  {Breitschwerdt}  \& {Dettmar}}{{T{\"u}llmann} et~al.}{2006}]{tullman:2006}
{T{\"u}llmann} R.,  {Pietsch} W.,  {Rossa} J.,  {Breitschwerdt} D.,   {Dettmar}
  R.~J.,  2006, \mn@doi [\aap] {10.1051/0004-6361:20052936}, \href
  {https://ui.adsabs.harvard.edu/\#abs/2006A&A...448...43T} {448, 43}

\bibitem[\protect\citeauthoryear{{Tumlinson}, {Peeples}  \& {Werk}}{{Tumlinson}
  et~al.}{2017}]{tumlinson:2017}
{Tumlinson} J.,  {Peeples} M.~S.,   {Werk} J.~K.,  2017, \mn@doi [\araa]
  {10.1146/annurev-astro-091916-055240}, \href
  {http://adsabs.harvard.edu/abs/2017ARA%26A..55..389T} {55, 389}

\bibitem[\protect\citeauthoryear{Van~Rossum \& Drake}{Van~Rossum \&
  Drake}{2009}]{python}
Van~Rossum G.,  Drake F.~L.,  2009, Python 3 Reference Manual.
CreateSpace, Scotts Valley, CA

\bibitem[\protect\citeauthoryear{{Vanderlinde} et~al.,}{{Vanderlinde}
  et~al.}{2010}]{vanderlinde:2010}
{Vanderlinde} K.,  et~al., 2010, \mn@doi [\apj] {10.1088/0004-637X/722/2/1180},
  \href {https://ui.adsabs.harvard.edu/abs/2010ApJ...722.1180V} {722, 1180}

\bibitem[\protect\citeauthoryear{{Vikhlinin}, {Kravtsov}, {Forman}, {Jones},
  {Markevitch}, {Murray}  \& {Van Speybroeck}}{{Vikhlinin}
  et~al.}{2006}]{vikhlinin:2006}
{Vikhlinin} A.,  {Kravtsov} A.,  {Forman} W.,  {Jones} C.,  {Markevitch} M.,
  {Murray} S.~S.,   {Van Speybroeck} L.,  2006, \mn@doi [\apj]
  {10.1086/500288}, \href
  {https://ui.adsabs.harvard.edu/abs/2006ApJ...640..691V} {640, 691}

\bibitem[\protect\citeauthoryear{{Wang}}{{Wang}}{2010}]{wang:2010}
{Wang} Q.~D.,  2010, \mn@doi [Proceedings of the National Academy of Science]
  {10.1073/pnas.0914255107}, \href
  {https://ui.adsabs.harvard.edu/abs/2010PNAS..107.7168W} {107, 7168}

\bibitem[\protect\citeauthoryear{{Wang}, {Whitaker}  \& {Williams}}{{Wang}
  et~al.}{2005}]{wang:2005}
{Wang} Q.~D.,  {Whitaker} K.~E.,   {Williams} R.,  2005, \mn@doi [\mnras]
  {10.1111/j.1365-2966.2005.09379.x}, \href
  {https://ui.adsabs.harvard.edu/abs/2005MNRAS.362.1065W} {362, 1065}

\bibitem[\protect\citeauthoryear{{Wang}, {Li}, {Jiang}  \& {Fang}}{{Wang}
  et~al.}{2016}]{wang:2016}
{Wang} Q.~D.,  {Li} J.,  {Jiang} X.,   {Fang} T.,  2016, \mn@doi [\mnras]
  {10.1093/mnras/stv2886}, \href
  {https://ui.adsabs.harvard.edu/abs/2016MNRAS.457.1385W} {457, 1385}

\bibitem[\protect\citeauthoryear{{White} \& {Frenk}}{{White} \&
  {Frenk}}{1991}]{white:1991}
{White} S.~D.~M.,  {Frenk} C.~S.,  1991, \mn@doi [\apj] {10.1086/170483}, \href
  {http://adsabs.harvard.edu/abs/1991ApJ...379...52W} {379, 52}

\bibitem[\protect\citeauthoryear{{White} \& {Rees}}{{White} \&
  {Rees}}{1978}]{white:1978}
{White} S.~D.~M.,  {Rees} M.~J.,  1978, \mn@doi [\mnras]
  {10.1093/mnras/183.3.341}, \href
  {http://adsabs.harvard.edu/abs/1978MNRAS.183..341W} {183, 341}

\bibitem[\protect\citeauthoryear{{White}, {Navarro}, {Evrard}  \&
  {Frenk}}{{White} et~al.}{1993}]{white:1993}
{White} S. D.~M.,  {Navarro} J.~F.,  {Evrard} A.~E.,   {Frenk} C.~S.,  1993,
  \mn@doi [\nat] {10.1038/366429a0}, \href
  {https://ui.adsabs.harvard.edu/abs/1993Natur.366..429W} {366, 429}

\bibitem[\protect\citeauthoryear{Wiersma, Schaye  \& Smith}{Wiersma
  et~al.}{2009a}]{wiersma:2009}
Wiersma R. P.~C.,  Schaye J.,   Smith B.~D.,  2009a, \mn@doi [Mon. Not. Roy.
  Astron. Soc.] {10.1111/j.1365-2966.2008.14191.x}, 393, 99

\bibitem[\protect\citeauthoryear{{Wiersma}, {Schaye}, {Theuns}, {Dalla Vecchia}
   \& {Tornatore}}{{Wiersma} et~al.}{2009b}]{wiersma:2009b}
{Wiersma} R. P.~C.,  {Schaye} J.,  {Theuns} T.,  {Dalla Vecchia} C.,
  {Tornatore} L.,  2009b, \mn@doi [\mnras] {10.1111/j.1365-2966.2009.15331.x},
  \href {https://ui.adsabs.harvard.edu/abs/2009MNRAS.399..574W} {399, 574}

\bibitem[\protect\citeauthoryear{{van de Voort}, {Quataert}, {Hopkins},
  {Faucher-Gigu{\`e}re}, {Feldmann}, {Kere{\v{s}}}, {Chan}  \& {Hafen}}{{van de
  Voort} et~al.}{2016}]{voort:2016}
{van de Voort} F.,  {Quataert} E.,  {Hopkins} P.~F.,  {Faucher-Gigu{\`e}re}
  C.-A.,  {Feldmann} R.,  {Kere{\v{s}}} D.,  {Chan} T.~K.,   {Hafen} Z.,  2016,
  \mn@doi [\mnras] {10.1093/mnras/stw2322}, \href
  {https://ui.adsabs.harvard.edu/abs/2016MNRAS.463.4533V} {463, 4533}

\bibitem[\protect\citeauthoryear{{van der Walt}, {Colbert}  \&
  {Varoquaux}}{{van der Walt} et~al.}{2011}]{numpy}
{van der Walt} S.,  {Colbert} S.~C.,   {Varoquaux} G.,  2011, Computing in
  Science Engineering, 13, 22

\makeatother
\end{thebibliography}

\appendix

\section{Hot gas self-similarity}\label{appendix:self-similar}

\begin{figure}
    \includegraphics[width=1.00\linewidth]{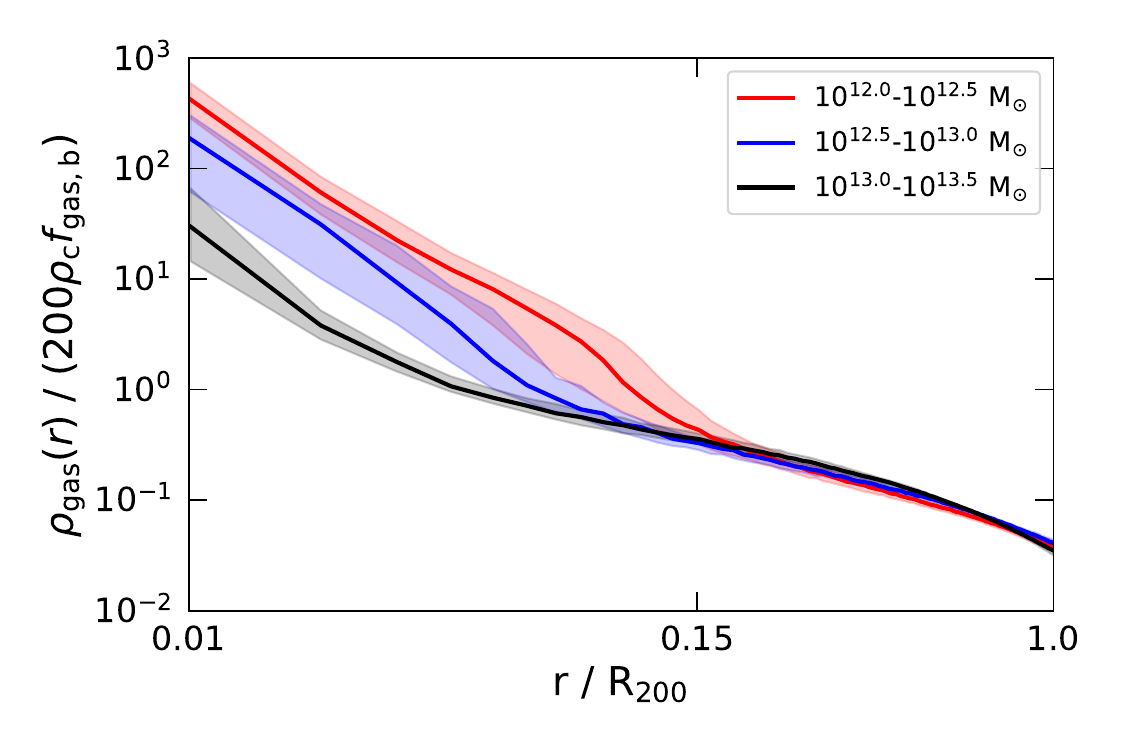} 
    \caption{Three-dimensional, spherically averaged radial density
      profiles of the gas within halos of disc galaxies in the EAGLE
      Ref-L100N1504 simulation.  We show the median gas density
      profile in halos in mass bins of: \protect
      $10^{12}-10^{12.5} \units{\msun}$ (red),
      $10^{12.5}-10^{13.0} \units{\msun}$ (blue) and
      $10^{13}-10^{13.5} \units{\msun}$ (black). The density
      profiles are normalised by $200$ times the critical mass density
      of the universe, the cosmic baryon-to-dark matter ratio and the
      gas fraction of each halo. The bands enclose the $30$th and
      $70$th percentiles.}
    \label{fig:self-similar}
\end{figure}

The radial profiles of the gas density estimated from all of the
galaxies in the EAGLE Ref-L100N1504 simulated in the halo mass range,
$10^{12}-10^{13.5} \units{\msun}$, are shown in Fig.~\ref{fig:self-similar}.
The profiles are normalised by the
individual baryon-to-dark-matter ratio within each halo. We see
that the normalised gas density profiles in the central region vary
significantly as a function of halo mass. However, in the outer
regions, the gas density profiles exhibit self-similarity across two
orders of magnitude of halo mass. This self-similarity also appears
to be present in the \textsc{GIMIC} simulations \citep{crain:2009}
as may be seen in the top-left panel of fig~8 of 
\cite{crain:2010}. While the slopes of the gas density profiles are
clearly self-similar in \cite{crain:2010}, the normalisation increases
with halo mass due to the increasing gas fraction in high-mass halos.

In the central regions, the normalised gas density, at a fixed radius,
is much higher in lower mass halos. This likely reflects the differing
impact of feedback in these halos. In lower mass halos AGN feedback
and, in some cases, SNe feedback, is able to eject gas to the virial
radius and beyond. This will lower the total gas fraction of the halo,
but if the process takes place over a long timescale, the density
profile should remain unaffected. In more massive halos, 
winds driven by AGN are unlikely to leave the halo, and thus the net
result is that gas is transported outwards. This process acts to
increase the density at large radii while decreasing it at small radii.

The self-similarity of the hot gas profiles in the outer regions of
the halo validates a key assumption in the derivation of
Eqn.~\ref{eq:lx_halo_mass_approx}. The observed self-similarity
reflects the fact that the gravitational forces are dominated by the
dark matter distribution, which has previously been shown to be well
converged for different subgrid physics models \citep{schaye:2015}. As
such, when appropriately normalised by their individual gas fractions,
the gas radial density profiles in the radial range,
$(0.15-1.00)~R_{500}$, agree in both trend and normalisation. This is
also expected from Fig.~\ref{fig:eagle_scaling_prist_wind} in which we
showed that the X-ray emission in this radial range (and the mass
fraction which is not shown) are dominated by accreted gas. Although
the gas density profiles in the EAGLE simulations display
self-similarity, it is not clear whether this behaviour will be also
be present in other cosmological hydrodynamical simulations or,
indeed, in the real universe. Interesting future work might assess how
well these results hold across other simulations with different
subgrid models and parameters.

\section{X-ray emission}\label{appendix:xray}

\begin{figure*}
  \includegraphics[width=1.00\textwidth]{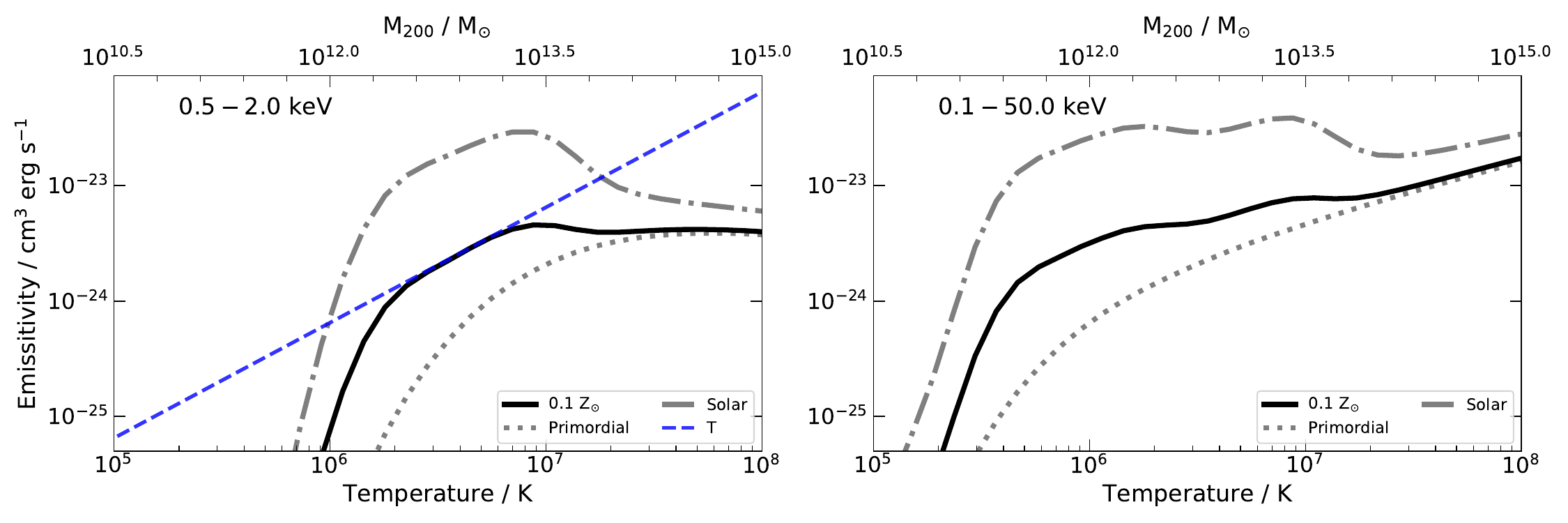}
  \caption{The black-solid, grey-dashed and grey-dotted lines show the
    cooling function in the $(0.5 - 2.0) \units{keV}$ (left) and
    $(0.1 - 50) \units{keV}$ (right) band from all ions, assuming a
    metallicity of $0.1 Z_{\odot}$, $Z_{\odot}$ and the primordial
    value \citep{anders:1989}, as a function of temperature using the
    AtomDB v3.0.1 code.  The approximation, $\Lambda \propto T$, to
    the soft X-ray cooling function is shown by the dashed-blue line
    for $Z = 0.1 Z_{\odot}$.}
    \label{fig:emiss}
\end{figure*}

In Fig.~\ref{fig:emiss} we plot the cooling function from all ions
assuming primordial abundance as well as metallicities $Z_{\odot}$,
$0.1 Z_{\odot}$ \citep{anders:1989}, as a function of temperature
using the AtomDB (v3.0.1) in the two energy bands
$(0.5- 50) \units{keV}$ and $(0.5 - 2.0) \units{keV}$. We also show an
approximation to the total soft X-ray cooling function for
$Z = 0.1 Z_{\odot}$. In general, we find that the behaviour of the
cooling function in the $(0.5 - 2.0) \units{keV}$ energy band can be
well described by a simple power law relation $\Lambda \propto T$ for
$T \approx (10^6-10^7) \units{K}$ and $\Lambda = \mathrm{const}$ for
$T \approx (10^7-10^8) \units{K}$.

These approximations do not adequately capture the rapid increase in
the cooling function at a temperature of $T \approx 10^{6}
\units{K}$. They also do not appropriately model the flattening of the
soft X-ray cooling function at temperatures above $10^{7} \units{K}$,
which is due to a significant fraction of the photons having energies
outside the selected range. In general, these inadequacies have little
impact in the halo mass range, $10^{12}-10^{13.5} \units{\msun}$, in
which we apply these approximations. The right panel of
Fig.~\ref{fig:emiss} shows that a power law can describe well the
cooling function in the $(0.1 - 50) \units{keV}$ band over an even
broader range of halo mass; however, there is limited observational
data in that energy band.

A variety of observational estimates of the metallicity of gaseous
coronae have yielded values of $Z \approx 0.1
Z_{\odot}$. This is consistent with the hot halo gas metallicity we
find in the EAGLE simulations and in previous hydrodynamic simulations
\citep[e.g.][]{crain:2010}. It may be seen in Fig.~\ref{fig:emiss}
that the slope of the cooling function is largely independent of the
metallicity which, however, has a larger effect on the
normalisation. Therefore the metallicity of the gaseous corona is
unimportant, as long as it does not vary significantly with halo mass.

\section{Stellar-halo mass relationship}\label{appendix:stellar_halo}

\begin{figure*}
    \includegraphics[width=1.00\textwidth]{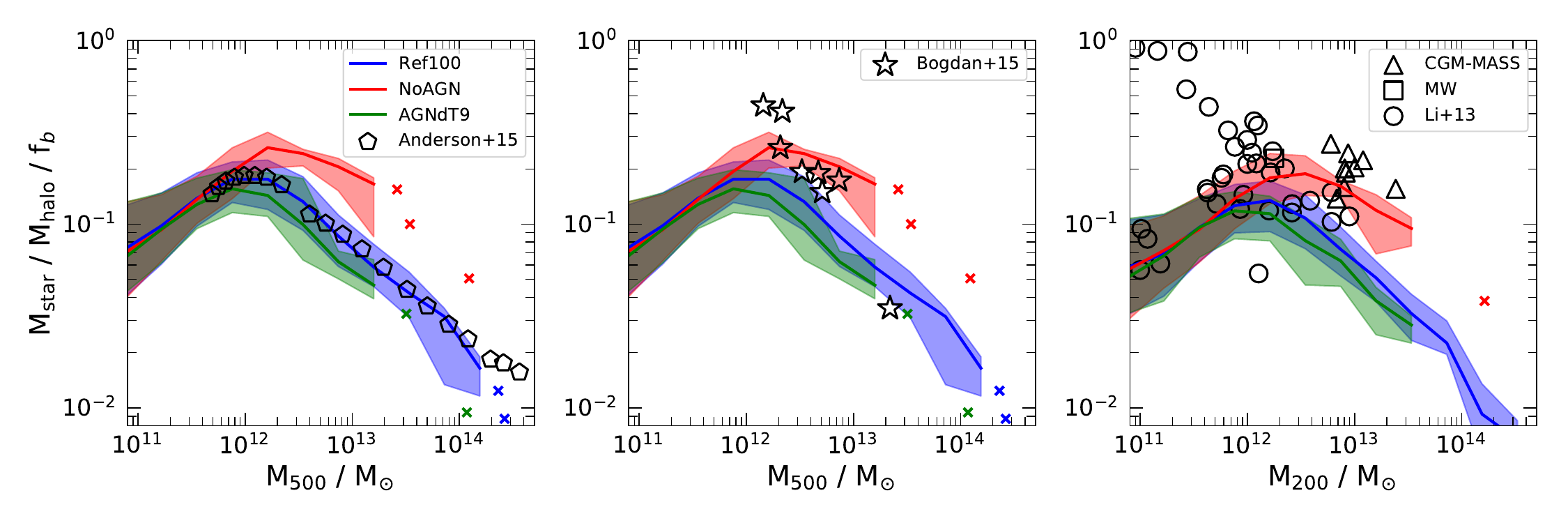} 
    \caption{The stellar-to-halo mass ratio, normalised by the mean
      universal baryon fraction, $f_{b}$, as a function of both
      $M_{500}$ (left panel) and $M_{200}$ (central and right
      panels). Results from three different EAGLE simulations,
      Ref-L100N1504 (blue), NoAGN-L050N0752 (red) and AGNdT9-L050N0752
      (green), are shown as medians in halo mass bins of
      $0.20 \protect\units{dex}$. When there are fewer than five
      objects in a bin, we plot the results of individual galaxies. We
      also show the 20th to 80th percentiles of the distribution as
      shaded regions. The observational results of
      \protect\cite{anderson:2015} (left), \protect\cite{bogdan:2015}
      (center) and \protect\cite{li:2017} (right) are plotted.}
    \label{fig:stellar_halo_fixed}
\end{figure*}

In Fig.~\ref{fig:stellar_halo_fixed} we plot the stellar-to-halo mass
ratio normalised by the mean universal baryon fraction, $f_{b}$, as a
function of $M_{500}$ (left panel) and $M_{200}$ (central and right
panels). The results from three different EAGLE simulations,
Ref-L100N1504, NoAGN-L050N0752 and AGNdT9-L050N0752, are shown. We
also plot the observational data of \cite{anderson:2015} (left),
\cite{bogdan:2015} (center) and \cite{li:2017} (right). We define the
stellar mass as the total mass of stars within a fixed spherical
aperture of radius $30 \units{pkpc}$. This aperture is chosen as
\cite{schaye:2015} found that it yields stellar masses similar to
those inferred from the Petrosian-r band aperture often used in
observational studies.


The two EAGLE simulations, Ref-L100N1504 and AGNdT9-L050N0752, have a
consistent stellar-to-halo mass relation which, moreover, agrees with
the results of \cite{moster:2013} from abundance matching
\citep{schaye:2015}.  However, the NoAGN-L050N0752 simulation
overpredicts the stellar-mass in galaxies of
$\geq 10^{12} \units{\msun}$. The stellar-halo mass relation of the
two main simulations, Ref-L100N1504 and AGNdT9-L050N0752, broadly
agree with the observations of \cite{anderson:2015} and
\cite{bogdan:2015}. However, the sample of \cite{li:2017} contains a
population of low mass halos with very high stellar masses which are
not found in the EAGLE simulations.

The stellar-dominated galaxies at low halo masses of \cite{li:2017}
are inconsistent with the abundance matching predictions of
\cite{moster:2013}. The cause of this could be an incorrect inference
of the halo mass from the measured rotation velocities. The
uncertainty in the halo masses complicates the interpretation of the
observational \lxmvir\ relation. 


\bsp
\label{lastpage}
\end{document}